\documentclass[12pt]{article}
\usepackage{amsmath}
\usepackage{graphicx}
\usepackage{natbib}
\usepackage{url} % not crucial - just used below for the URL 

\usepackage[utf8]{inputenc} % allow utf-8 input
\usepackage[T1]{fontenc}    % use 8-bit T1 fonts
\usepackage{hyperref}       % hyperlinks
\usepackage{url}            % simple URL typesetting
\usepackage{booktabs}       % professional-quality tables
\usepackage{amsfonts}       % blackboard math symbols
\usepackage{nicefrac}       % compact symbols for 1/2, etc.
\usepackage{microtype}      % microtypography
\usepackage{xcolor}         % colors
\usepackage{setspace}
\usepackage{graphicx}
\usepackage{subfigure}
\usepackage{adjustbox}
\usepackage{paralist}

\usepackage{amsmath}
\usepackage{amssymb}
\usepackage{mathtools}
\usepackage{amsthm}
\usepackage{bbm}
\allowdisplaybreaks

\usepackage{algorithm}
\usepackage{algorithmic}

\theoremstyle{plain}
\newtheorem{theorem}{Theorem}[section]
\newtheorem{proposition}{Proposition}
\newtheorem{lemma}{Lemma}
\newtheorem{corollary}{Corollary}
\newtheorem{definition}{Definition}
\newtheorem{assumption}{Assumption}
\newtheorem*{remark}{Remark}

\usepackage[textsize=tiny]{todonotes}
\usepackage{enumitem}
\newcommand{\indep}{\perp \!\!\! \perp}
\newcommand{\E}{\mathrm{E}}

\newcommand{\pr}{\mathrm{pr}}

\DeclareMathOperator*{\argmin}{arg\,min}
\allowdisplaybreaks

\DeclareMathOperator{\Lap}{Lap}
\DeclareMathOperator{\clip}{clip}

\renewenvironment{enumerate}[1]{\begin{compactenum}#1}{\end{compactenum}}

\usepackage{bibspacing}
\newcommand{\blind}{0}

\begin{document}

\def\spacingset#1{\renewcommand{\baselinestretch}%
{#1}\small\normalsize} \spacingset{1}

%%%%%%%%%%%%%%%%%%%%%%%%%%%%%%%%%%%%%%%%%%%%%%%%%%%%%%%%%%%%%%%%%%%%%%%%%%%%%%

\if0\blind
{
  \title{\bf Differentially Private Covariate Balancing Causal Inference}
  \author{Yuki Ohnishi\hspace{.2cm}\\
    Department of Statistics, University of Florida\\
    and \\
    Jordan Awan\thanks{
    This work was supported in part by the National Science Foundation (NSF) grants SES
2150615.}\\
    Department of Statistics, University of Pittsburgh}
  \maketitle
} \fi

\if1\blind
{
  \bigskip
  \bigskip
  \bigskip
  \begin{center}
    {\LARGE\bf Differentially Private Covariate Balancing Causal Inference}
\end{center}
  \medskip
} \fi

\bigskip
\begin{abstract}
% Differential privacy is the leading mathematical framework for privacy protection, providing a probabilistic guarantee that safeguards individuals' private information when publishing statistics from a dataset. For causal inference from observational data, a central challenge is maintaining covariate balance between treatment groups while protecting both balance diagnostics and treatment-effect estimates. In this article, we present a differentially private two-stage covariate balancing weighting estimator to infer causal effects from observational data. Our algorithm produces both point and interval estimators with statistical guarantees, such as consistency and rate optimality, under a given privacy budget. 
Differential privacy is the leading mathematical framework for privacy protection, providing a probabilistic guarantee that safeguards individuals' private information when publishing statistics from a dataset. This guarantee is achieved by applying a randomized algorithm to the original data, which introduces unique challenges in data analysis by distorting inherent patterns. In particular, causal inference using observational data while preserving privacy is challenging because it requires a good covariate balance between treatment groups, but checking the covariate balance is challenging when privacy is a primary concern because releasing any statistic (e.g., t-statistic) for the balance check compromises privacy. Additionally, the performance of the privatized estimator critically depends on the choice of differential privacy mechanisms, and it remains unexplored how privacy-protecting causal inference could be conducted with observational data. In this article, we present a differentially private two-stage covariate balancing weighting estimator to infer causal effects from observational data. Our algorithm produces both point and interval estimators with statistical guarantees, such as covariate balance, rate optimality, and oracle budget allocation, under a given privacy budget. 
\end{abstract}

\section{Introduction}
Research, policy-making, and business increasingly rely on sensitive data, creating a need for methods that balance data accessibility with privacy protection.
Differential privacy (DP) has emerged as the gold standard for achieving this balance. It is increasingly adopted in industry \citep{erlingsson2014, apple2017} and by government agencies, such as the U.S. Census Bureau \citep{census2018}. DP offers a probabilistic guarantee of privacy, safeguarding against arbitrary breaches by applying a privacy mechanism (e.g., adding calibrated random noise) to summary statistics and synthetic data before their release to the public. However, while DP effectively protects privacy, integrating it into statistical analyses introduces significant challenges.

In particular, causal inference is essential for decision-making across various fields. Randomized experiments are ideal for identifying causal effects, but their costs and complexities often necessitate the use of observational data. However, observational studies face challenges, particularly in adjusting for confounding variables, which can bias treatment effect estimates. The propensity score \citep{ROSENBAUM1086} helps eliminate confounding bias when treatment relies on observables and has been adopted in many applications like matching, stratification, and weighting \citep{Rosenbaum_Rubin_1985, Abadie_Imbens_2006, Heckman1997}.
\citet{ROSENBAUM1086} demonstrated that the true propensity score mitigates the confounding bias by creating quasi-randomized experiments in which covariate distributions are well-balanced between treatment groups. However, accurately specifying the true propensity score model is typically difficult, and even slight misspecifications can introduce significant bias \citep{Kang_Schafer_2007}.

To ensure robust analyses under the misspecification of the propensity score, the focus of the propensity score analysis literature has shifted from accurately predicting treatment assignment to instead using it to achieve a good covariate balance between treatment groups \citep{Stuart2010, Imai2014, imbens_rubin_2015}. 
Balance diagnostics are therefore essential, but releasing them from confidential data requires privacy protection. Repeated checks during model refinement consume additional privacy budget.
A promising approach is the covariate balancing weighting scheme
\citep{Hainmueller2012,Chuen2015,Zhao2017,Zhao2019,Hazlett2020,Kong0223,Jianqing2023,Huling2024}. These methods ensure automatic covariate balance by solving optimization problems for propensity score estimation under empirical covariate balancing constraints. Our focus is to retain quantitative control of this balance while privatizing both the fitted weights and the normalized causal contrast.

Our contribution is several-fold. First, we propose a two-stage private covariate-balancing estimator using the covariate balancing scoring rules (CBSR) of \citet{Zhao2019} and the K-norm gradient (KNG) mechanism \citep{Reimherr2019}. KNG samples a propensity parameter according to the aggregate score norm. Four Laplace releases protect the weighted outcomes and denominators.

Second, we derive theoretical guarantees of our estimator for covariate balance, estimation, and inference under privacy protection. Explicit finite-sample bounds control covariate imbalance and the discrepancy between the private and nonprivate estimators, retaining privacy budgets. We also show our estimator matches the population-risk bounds, establishing consistency and minimax-rate optimality. Following \citet{barber2014,Cai2021}, optimality concerns total risk jointly in sample size $n$ and privacy budget $\epsilon$, with constants independent of both and uniform over the model class. In particular, the privacy budget sequence $\epsilon_n=n^{-1/2}$ preserves the $n^{-1}$ mean squared error rate. In this setting, we also derive the joint sampling and privacy law, identify the oracle allocation across propensity fitting and the four queries, and construct confidence intervals using privately estimated moments. These intervals have pointwise asymptotic coverage even when privacy noise contributes at first order. Section~\ref{sec:growing-series} further gives explicit conditions under which privatization preserves independently established nonprivate sampling limits for growing bases.

Finally, simulation studies evaluate estimation error, covariate balance, and interval coverage, including comparisons with existing private propensity-weighting methods and an assessment of the oracle budget allocation. We also analyze the publicly available email campaign data of \citet{Hillstrom2008Public} to examine covariate balancing under privacy protection. The private estimates of the email's effect on website visits closely reproduce the nonprivate CBSR estimate. Mean covariate imbalance decreases as the privacy budget increases, and interval widths approach their nonprivate counterpart at larger budgets. This application illustrates how privacy protection affects balance and covariate-adjusted point and interval estimates.

Section~\ref{sec:preliminaries} introduces the framework, Section~\ref{sec:methodologies} gives the method and guarantees, and Sections~\ref{sec:simulation} and \ref{sec:analysis} present the empirical studies. Proofs and additional diagnostics are in the supplement.

\subsection{Current Literature}
Although DP is a rapidly expanding field, research focusing on propensity score analyses for observational data in privacy-sensitive contexts, similar to ours, remains limited.
\citet{Lee2019} proposed a privacy-preserving inverse propensity score estimator for estimating the average treatment effect (ATE). They suggested a Horvitz-Thompson-type estimator using an objective perturbation technique to ensure privacy. However, this approach requires regularization, resulting in a residual bias in the propensity score, even asymptotically. Additionally, they did not consider private confidence intervals or explore covariate balance, which is central to causal inference from observational data.
\citet{guha2024} developed a causal inference methodology that leverages the subsample-and-aggregate algorithm \citep{Nissim2007} to estimate the weighted average treatment effects with binary outcomes. They also presented private standard errors and confidence intervals for the estimator. However, they did not consider the covariate balance that we consider in this paper, and the performance guarantees of their estimator depend on the number of subgroups, which is a hyperparameter for the algorithm that analysts must tune for the application at hand.

Some authors have addressed causal inference problems under different DP paradigms or studied designs for causal inference than those used in our work.
\citet{D'Orazio2015} introduced differential privacy mechanisms in causal inference and algorithms for releasing private estimates of causal effects, mainly for experimental settings. \citet{Kusner2016PrivateCI} explored causal inference using the additive noise model, a more restrictive approach than the potential outcomes framework considered in this paper.
\citet{Komarova2020} demonstrated that under differential privacy, identifying causal parameters fails in regression discontinuity designs. \citet{Agarwal2021} studied causal inference with corrupted data, including privacy mechanisms. \citet{ohnishi2025} developed inference for randomized experiments under local DP, while \citet{niu2022} studied private heterogeneous treatment effect estimation.
\citet{Javanmard2024} proposed Cluster-DP for randomized trials, while \citet{Chen2024} studied the experimental design problem under Distributed DP with secure aggregation.

\section{Preliminaries}
\label{sec:preliminaries}
\subsection{Notation and Causal Estimands}
\label{sec:setting}
Throughout this manuscript, we adopt the Rubin Causal Model \citep{imbens_rubin_2015} as our causal inference framework.  We consider $n$ units, indexed by $i = 1, \ldots , n$,  as a random sample from a large super-population. Each unit $i$ has an outcome $Y_i  \in [0,1]$, treatment assignment $Z_i \in \{0,1\}$, and covariates $X_i \in \mathcal{X}$, respectively. 
The covariate space $\mathcal X$ is public. We assume no interference or hidden versions of treatment and write $Y_i(z)\in[0,1]$ for the potential outcome under $z\in\{0,1\}$, with observed record $O_i=(X_i,Z_i,Y_i)$ and $Y_i=Z_iY_i(1)+(1-Z_i)Y_i(0)$. The records are iid and the sample size $n$ is public. For a publicly specified outcome range $[a_Y,b_Y]$, a fixed affine rescaling gives this setup. Clipping an unbounded outcome changes the causal target.
\begin{assumption}[Identification]
\label{asmp:identification}\label{asmp:positivity}\label{asmp:unconfoundedness}
For a public constant $0<\eta<1/2$, the true propensity $e_0(x)=P(Z=1\mid X=x)$ satisfies $\eta\leq e_0(x)\leq1-\eta$ on $\mathcal X$, and $\{Y(0),Y(1)\}\indep Z\mid X$.
\end{assumption}

We define the average treatment effect (ATE) as $\tau = \E\{Y(1)-Y(0)\}$.
The ATE summarizes effects across the whole population. Other targets may be more relevant when an intervention applies only to selected units, the observed sample differs from the intended population, or overlap is limited.
\citet{Li2018} consider balancing weights for a target population whose density is proportional to $f(x)h_0(x)$, where $f$ is the covariate density with respect to a dominating measure. Equivalently, we define
\begin{equation}
\label{eq:wate}
\tau_h=\frac{\E[h_0(X)\{Y(1)-Y(0)\}]}{\E h_0(X)}.
\end{equation}
The choices $h_0=1$, $e_0$, $1-e_0$, and $e_0(1-e_0)$ give the ATE, ATT, ATC, and average treatment effect in the overlap population (ATO). The functional form of $h_0$ is chosen by the analyst, but its values generally depend on the unknown propensity.

\subsection{Covariate Balancing Propensity Score Estimation}
\label{sec:cbsr}
Generally, propensity score analysis can be viewed as a decision problem through the lens of statistical decision theory.
Specifically, for binary treatment, we consider a scoring rule $S(e,z)$ which evaluates a predicted assignment probability $e\in[0,1]$ after observing $z\in\{0,1\}$. A rule is proper if, under any Bernoulli distribution, its expected score is maximized by the true assignment probability, and strictly proper if this maximizer is unique \citep{Gneiting2007}.
For $0<e<1$, the representation of \citet{Savage1971} writes a differentiable proper scoring rule as $S(e,1)=G(e)+(1-e)G'(e)$ and $S(e,0)=G(e)-eG'(e)$, where $G:(0,1)\to\mathbb R$ is convex. If $G$ is twice differentiable, its score derivative satisfies
\begin{equation}
\label{eq:score_rule_deriv_p}
    \frac{\partial}{\partial e} S(e, z) = (z - e)G''(e) \text{ for } z = 0, 1.
\end{equation}
While many choices exist for $G$, we consider the Beta family for the class of proper scoring rules:
\begin{equation}
\label{eq:beta_family}
    G''_{\alpha, \beta}(e) = e^{\alpha - 1}(1 - e)^{\beta - 1}, \quad -\infty < \alpha, \beta < \infty.
\end{equation}

\citet{Zhao2019} proposed the covariate balancing scoring rule (CBSR) using the Beta family \eqref{eq:beta_family}, interpreting the estimating equation as the first-order covariate balancing constraint.
We use the logistic model $e_\theta(x)=\{1+\exp[-\theta^\top\phi(x)]\}^{-1}$, where $\phi:\mathcal X\to\mathbb R^d$ is a public feature map. The main minimax and inference results concern fixed $d$. A growing series basis is allowed in Corollary~\ref{cor:privacy_transfer} in Section~\ref{sec:growing-series}, where both the envelope and the empirical curvature are tracked explicitly.
\begin{assumption}[Public analysis domain]
\label{asmp:public}
The public feature map has envelope $C_\phi=\sup_x\|\phi(x)\|_2$ with $0<C_\phi<\infty$. The parameter set $\Theta\subset\mathbb R^d$ is compact and convex with positive volume, is chosen independently of the database, and ensures $e_\theta(x)\in[\eta,1-\eta]$ for every $x\in\mathcal X$ and $\theta\in\Theta$.
\end{assumption}
For example, a Euclidean ball centered at zero with radius $\log\{(1-\eta)/\eta\}/C_\phi$ is valid. Neither a unit-ball nor a Cartesian-product restriction on the raw covariates is needed. 

% For $\alpha,\beta\in[-1,0]$, $L_n$ is the average negative score, with a fixed measurable rule for ties in its minimization,
% \begin{align}
% \label{eq:loss}
% L_n(\theta)&=-\frac1n\sum_{i=1}^n S_{\alpha,\beta}\{e_\theta(X_i),Z_i\},
% \qquad \hat\theta\in\argmin_{\theta\in\Theta}L_n(\theta),\\
% \label{eq:gradient}
% g_n(\theta)&=\nabla L_n(\theta)=-\frac1n\sum_{i=1}^n s_i(\theta),\\
% s_i(\theta)&=\{Z_i-e_\theta(X_i)\}e_\theta(X_i)^\alpha\{1-e_\theta(X_i)\}^\beta\phi(X_i).\nonumber
% \end{align}
% \citet{Zhao2019} showed that the arm weights for target estimands are given by $w_1(e)=e^\alpha(1-e)^{\beta+1}$ and $w_0(e)=e^{\alpha+1}(1-e)^\beta$. Then $s_i=\{Z_iw_1(e_i)-(1-Z_i)w_0(e_i)\}\phi(X_i)$. An interior score root therefore balances the empirical weighted feature sums. A constrained boundary minimizer need not have zero gradient.

% The corresponding tilt is $h_0(x)=e_0(x)^{\alpha+1}\{1-e_0(x)\}^{\beta+1}$. Thus $\tau_h$ in \eqref{eq:wate} is the target $\tau_{\alpha,\beta}$ of the chosen scoring rule. With
% \[
%  A_z(\theta)=\frac1n\sum_i1\{Z_i=z\}w_z\{e_\theta(X_i)\}Y_i,
%  \qquad
%  B_z(\theta)=\frac1n\sum_i1\{Z_i=z\}w_z\{e_\theta(X_i)\},
% \]
% the nonprivate H\'ajek estimator is $A_1(\hat\theta)/B_1(\hat\theta)-A_0(\hat\theta)/B_0(\hat\theta)$ when its denominators are positive. Section~\ref{sec:methodologies} defines its protected version on every database, including an empty treatment arm.

For $\alpha,\beta\in[-1,0]$, define the average negative score and its
minimizer by
\begin{equation}
\label{eq:loss}
L_n(\theta)
=-\frac1n\sum_{i=1}^n S_{\alpha,\beta}\{e_\theta(X_i),Z_i\},
\qquad
\hat\theta\in\argmin_{\theta\in\Theta}L_n(\theta),
\end{equation}
with ties resolved by a fixed measurable rule. The logistic model gives
$\nabla_\theta e_\theta(x)
=e_\theta(x)\{1-e_\theta(x)\}\phi(x)$.
Combining this identity with \eqref{eq:score_rule_deriv_p} and
\eqref{eq:beta_family}, we obtain
\begin{align}
\label{eq:gradient}
g_n(\theta)
&=\nabla L_n(\theta)
=-\frac1n\sum_{i=1}^n s_i(\theta),\\
s_i(\theta)
&=\{Z_i-e_\theta(X_i)\}
  e_\theta(X_i)^\alpha
  \{1-e_\theta(X_i)\}^\beta\phi(X_i).
\nonumber
\end{align}

The connection to covariate balancing follows by separating the treated
and control contributions to the score. Writing $e_i=e_\theta(X_i)$,
the identity $Z_i-e_i=Z_i(1-e_i)-(1-Z_i)e_i$ yields $s_i(\theta) =\{Z_iw_1(e_i)-(1-Z_i)w_0(e_i)\}\phi(X_i),$
where $w_1(e)=e^\alpha(1-e)^{\beta+1}$ and $w_0(e)=e^{\alpha+1}(1-e)^\beta.$
These functions are the weights multiplying the treated and control feature vectors in the score equation. In particular, an interior minimizer satisfies $g_n(\hat\theta)=0$, and hence
\[
\sum_{i=1}^n Z_iw_1\{e_{\hat\theta}(X_i)\}\phi(X_i)
=
\sum_{i=1}^n(1-Z_i)w_0\{e_{\hat\theta}(X_i)\}\phi(X_i).
\]
Thus, fitting the propensity score with the chosen scoring rule balances
the empirical weighted sums of the supplied features
\citep{Zhao2019}. If the intercept belongs to the basis span, this
equation also equates the total weights in the two arms, giving equal
weighted feature means when the denominators are positive. A constrained
boundary minimizer need not have zero gradient.

These arm weights also determine the target population in
\eqref{eq:wate}. Before weighting, the covariate densities in the treated
and control groups are proportional to $f(x)e_0(x)$ and
$f(x)\{1-e_0(x)\}$, respectively. At the true propensity score, the
weights satisfy
\[
e_0(x)w_1\{e_0(x)\}
=
\{1-e_0(x)\}w_0\{e_0(x)\}
=
e_0(x)^{\alpha+1}\{1-e_0(x)\}^{\beta+1}
=:h_0(x).
\]
Consequently, weighting each arm by its corresponding $w_z$ produces
the same target covariate density, proportional to $f(x)h_0(x)$.
Under Assumption~\ref{asmp:identification}, the corresponding weighted
outcome means satisfy
\begin{align}
\label{eq:weighted_outcome_mean}
    \frac{\E[1\{Z=z\}w_z\{e_0(X)\}Y]}
     {\E[1\{Z=z\}w_z\{e_0(X)\}]} 
     = \frac{\E\{h_0(X)Y(z)\}}{\E h_0(X)}, 
     \qquad z\in\{0,1\}.
\end{align}
Their difference is therefore the WATE $\tau_h$ in \eqref{eq:wate},
which we denote by $\tau_{\alpha,\beta}$ for the chosen scoring rule.
For example, $(\alpha,\beta)=(-1,-1)$ gives inverse probability
weights and $h_0=1$, corresponding to the ATE, whereas
$(\alpha,\beta)=(0,0)$ gives overlap weights and $h_0=e_0(1-e_0)$,
corresponding to the ATO. Table~\ref{tab:estimands} summarizes the
common choices. The analyst selects the desired estimand within this
family and uses the corresponding $(\alpha,\beta)$ to specify the
scoring rule for propensity estimation.

Replacing the population expectations by empirical averages and
$e_0$ by the fitted propensity $e_{\hat\theta}$ gives the nonprivate
H\'ajek estimator. Specifically, let
\[
 A_z(\theta)=\frac1n\sum_i1\{Z_i=z\}w_z\{e_\theta(X_i)\}Y_i,
 \qquad
 B_z(\theta)=\frac1n\sum_i1\{Z_i=z\}w_z\{e_\theta(X_i)\},
\]
The nonprivate H\'ajek estimator is $A_1(\hat\theta)/B_1(\hat\theta)-A_0(\hat\theta)/B_0(\hat\theta)$ when its denominators are positive.

\begin{table}[t]
\centering
\caption{Targets, weights, and scoring rules. Scores are specified up to constants independent of $e$. $M$ is a public bound on either arm's weight defined in \eqref{eq:weight-bounds}.}
\label{tab:estimands}
\begin{tabular}{ccccccc}
\toprule
$(\alpha,\beta)$ & Target & $w_1(e)$ & $w_0(e)$ & $S(e,1) $ & $ S(e,0)$ & $M$\\
\midrule
$(-1,-1)$ & ATE & $1/e$ & $1/(1-e)$ & $\log\frac e{1-e}-\frac1e $ & $ \log\frac{1-e}e-\frac1{1-e}$ & $1/\eta$\\[3pt]
$(-1,0)$ & ATC & $(1-e)/e$ & $1$ & $-1/e $ & $ \log\frac{1-e}e$ & $(1-\eta)/\eta$\\[3pt]
$(0,-1)$ & ATT & $1$ & $e/(1-e)$ & $\log\frac e{1-e} $ & $ -1/(1-e)$ & $(1-\eta)/\eta$\\[3pt]
$(0,0)$ & ATO & $1-e$ & $e$ & $\log e $ & $ \log(1-e)$ & $1-\eta$\\
\bottomrule
\end{tabular}
\end{table}

For $\alpha,\beta\in[-1,0]$, define
\begin{equation}
\label{eq:weight-bounds}
M=\max_{z\in\{0,1\}}\sup_{e\in[\eta,1-\eta]}w_z(e),
\qquad
m=\min_{z\in\{0,1\}}\inf_{e\in[\eta,1-\eta]}w_z(e)>0.
\end{equation}
In particular, $M\leq1/\eta$. For $t=\theta^\top\phi(x)$ and
$e=e_\theta(x)$, the log-weight derivatives are
\begin{equation}
\label{eq:logweight-derivative}
k_1(e)=\frac{d\log w_1(e)}{dt}
=\alpha(1-e)-(\beta+1)e,
\qquad
k_0(e)=\frac{d\log w_0(e)}{dt}
=(\alpha+1)(1-e)-\beta e.
\end{equation}
They satisfy $-1\leq k_1(e)\leq0$ and $0\leq k_0(e)\leq1$, which imply that $L_n$ is convex in $\theta$.

\subsection{Differential Privacy}
\label{sec:dp}

We consider the central DP model, which involves a trusted data curator collecting and storing sensitive data from individuals in a central location. The data curator then performs a privacy-preserving analysis of the data and releases the results to the public.
Let $\mathcal{D}_n$ denote the collection of databases with $n$ units.
Let $\mathcal M$ be a randomized algorithm with input $D\in\mathcal D_n$. We say $ D, D' \in \mathcal{D}_n $ are adjacent if $d_{\mathrm{Ham}}(D,D')=1$, where $d_{\mathrm{Ham}}(D,D')$ is the Hamming distance between $D$ and $D'$.
% there exists only one record $ \{d\} \in D $ and one record $ \{d'\} \in D' $ such that $ d \neq d' $ and $ D - \{d\} = D' - \{d'\} $.
\begin{definition}[$ \epsilon $-Differential Privacy]
An algorithm $ \mathcal{M} $ satisfies $ \epsilon $-differential privacy ($ \epsilon $-DP), if for any pair of adjacent databases $ D,D' \in \mathcal{D}_n $, and any measurable set $ \mathcal A \subseteq \text{range}(\mathcal{M}) $, 
$
\pr\{\mathcal{M}(D) \in\mathcal A\} \leq \exp(\epsilon) \pr\{\mathcal{M}(D') \in\mathcal A\}.
$
\end{definition}
The privacy budget $\epsilon$ bounds how much replacing one record can change the output distribution. Smaller values give stronger privacy protection.

Two important properties of differential privacy are composition and invariance to post-processing \citep{Dwork2014}. Composition allows one to derive the cumulative privacy cost when releasing the results of multiple privacy mechanisms: if $\mathcal M_1$ is $\epsilon_1$-DP and $\mathcal M_2$ is $\epsilon_2$-DP, then the joint release $(\mathcal M_1(D),\mathcal M_2(D))$ satisfies $(\epsilon_1+\epsilon_2)$-DP. Invariance to post-processing ensures that applying a data-independent procedure to the output of a DP mechanism does not compromise the privacy guarantee: if $\mathcal M$ is $\epsilon$-DP with range $\mathcal Y$, and $f:\mathcal Y\rightarrow \mathcal Z$ is a (potentially randomized) function, then $f\circ \mathcal M$ is also $\epsilon$-DP. 

Noise calibration depends on sensitivity, the largest change in a statistic between adjacent databases. For $f:\mathcal D_n\to\mathbb R^k$, its $\ell_1$-sensitivity is
\[
 \Delta_f=\sup_{\substack{D,D'\in\mathcal D_n\\d_{\mathrm{Ham}}(D,D')=1}}
 \|f(D)-f(D')\|_1.
\]

\begin{proposition}[Laplace Mechanism]
\label{def:lapmech}
Let $\epsilon>0$ and $f:\mathcal D_n\to\mathbb R^k$ have finite sensitivity $\Delta_f$. If $\Delta_f>0$, the Laplace mechanism is
$\mathcal M(D)=f(D)+(\nu_1,\ldots,\nu_k)^\top$, where the $\nu_i$ are independent with
$\nu_i\sim\operatorname{Lap}(0,\Delta_f/\epsilon)$. The distribution
$\operatorname{Lap}(\mu,\varsigma)$ has density
$(2\varsigma)^{-1}\exp\{-|v-\mu|/\varsigma\}$
for $\varsigma>0$. Then $\mathcal M$ satisfies $\epsilon$-DP. If $\Delta_f=0$, releasing $f(D)$ without noise satisfies $\epsilon$-DP.
\end{proposition}

The K-Norm Gradient Mechanism (KNG) of \citet{Reimherr2019} privatizes parameters defined through optimization.

\begin{proposition}[K-Norm Gradient Mechanism (KNG)]
\label{def:KNG}
Let $\epsilon>0$, let $\Theta\subset\mathbb R^d$ be convex, and let $\|\cdot\|_K$ be a norm on $\mathbb R^d$. Suppose the measurable losses $\ell_n(\cdot;D):\Theta\to\mathbb R$ have gradients defined almost everywhere. A public measurable bound $\Delta:\Theta\to(0,\infty)$ satisfies
$\|\nabla\ell_n(\theta;D)-\nabla\ell_n(\theta;D')\|_K\leq\Delta(\theta)$
for every adjacent pair $D,D'\in\mathcal D_n$ and almost every $\theta\in\Theta$. Assume, for every $D\in\mathcal D_n$,
\[
0<\int_\Theta\exp\left\{-\frac{\epsilon}{2\Delta(\theta)}\|\nabla\ell_n(\theta;D)\|_K\right\}\,d\theta<\infty.
\]
Then sampling from the density with respect to Lebesgue measure on $\Theta$,
\begin{equation}
\label{eq:KNG_density}
 f_D(\theta)\propto\exp\left\{-\frac{\epsilon}{2\Delta(\theta)}\|\nabla\ell_n(\theta;D)\|_K\right\},
\end{equation}
satisfies $\epsilon$-DP.
\end{proposition}

\noindent
The KNG method starts with an objective function and favors summaries that nearly minimize it by weighting according to how close the gradient is to zero. Under certain technical conditions \citep[Theorem 3.2]{Reimherr2019}, the KNG achieves an asymptotic error of $O_p(n^{-1})$, which is asymptotically negligible compared to the statistical error for many problems. In contrast, under the same conditions, the exponential mechanism introduces noise of magnitude $O_p(n^{-1/2})$ \citep{Awan2019}. Furthermore, unlike objective perturbation \citep{Chaudhuri2011}, KNG does not require regularization.

\begin{remark}[Choice of privacy mechanism]
The exponential mechanism \citep{McSherry2007} provides a flexible approach to private optimization by sampling $\tilde\theta$ from $f_D(\theta)\propto\exp\{-c_0\ell_n(\theta;D)\}$, where $c_0$ is calibrated to the sensitivity of the loss and the privacy budget. Under suitable regularity conditions and a fixed privacy budget, its perturbation is of order $n^{-1/2}$ and is therefore not asymptotically negligible relative to root-$n$ sampling error \citep{Awan2019}.

Output and objective perturbation provide alternatives whose standard formulations use regularization to ensure strong convexity. For example, \citet{Lee2019} use Gaussian output perturbation to privatize a regularized propensity score estimator. For CBSR, regularization modifies the score equations that encode covariate balance and generally prevents exact balance. A penalty that does not vanish asymptotically can also induce persistent bias. Selecting the regularization parameter using confidential data requires accounting for this selection in the privacy guarantee. These considerations motivate applying KNG directly to the unregularized CBSR loss, whose gradient provides a direct measure of covariate imbalance.
\end{remark}

\section{Methodologies}
\label{sec:methodologies}
\subsection{Overview of Differentially Private Covariate Balancing Estimator}
\label{sec:private-estimator}
This section develops a two-stage private estimator of the WATE in \eqref{eq:wate}. The main difficulty in directly privatizing the fitted H\'ajek estimator is that its weights are estimated from the entire database. Specifically, the weights $w_z\{e_{\hat\theta(D)}(X_i)\}$ share a propensity parameter $\hat\theta(D)$ obtained by optimizing the CBSR loss in \eqref{eq:loss} using all $n$ records in $D$. Even when adjacent databases $D$ and $D'$ differ in only one record, refitting can change the propensity parameter and hence the fitted weights across all $n$ units, including units whose own records are unchanged. Averaging over $n$ records alone does not give a sensitivity bound proportional to $1/n$.

The calibration of noise using only the estimator's range $[-1,1]$ gives Laplace scale $2/\epsilon$, which does not decrease with sample size at a fixed privacy budget. \citet{guha2024} instead privatize an aggregate of estimates from disjoint subsets, with efficiency depending on the prespecified partition count.
To handle this dependence of all fitted weights on the database, our procedure consists of two steps. First we privatize the shared propensity parameter using KNG. The resulting weight summaries still use confidential covariates, treatment assignments, and outcomes, which require further protection. For every fixed first-stage output $\tilde\theta$, however, each of the four weighted averages $(A_1,B_1,A_0,B_0)$ has sensitivity $M/n$. The second stage therefore adds independent Laplace noise with scale $M/(n\epsilon_j)$ to each average and constructs the protected ratios. Adaptive composition accounts for both stages on the same database, giving point budget $\epsilon=\epsilon_\theta+\sum_{j=1}^4\epsilon_j$.

\subsection{Private Point Estimator}
\subsubsection{Private Propensity Score Estimation}
\label{sec:private-propensity}\label{sec:balance-main}
We first consider estimating the propensity score in a private manner via the KNG mechanism with $\lVert \cdot \rVert_K=\lVert \cdot \rVert_2$. 
Given budget $\epsilon_\theta>0$, the first stage samples from
\begin{equation}
\label{eq:kng}
\begin{split}
f_D(\theta)=\frac{\exp\{-a\|g_n(\theta;D)\|_2\}}{\int_\Theta\exp\{-a\|g_n(u;D)\|_2\}\,du}1\{\theta\in\Theta\},\qquad
a=\frac{n\epsilon_\theta}{4C_\phi M},
\end{split}
\end{equation}
For sampling, we adopt the privacy-aware rejection sampler \citep{Awan2024}, detailed in Appendix \ref{app:sampling}. We denote the KNG draw from this density by $\tilde\theta$.  The following lemma gives the gradient sensitivity required for the sampler and the required privacy guarantee.

\begin{lemma}[Public gradient sensitivity]
\label{lemma:sensitivity}
Under Assumption~\ref{asmp:public}, for every $\theta\in\Theta$ and adjacent $D,D'  \in \mathcal{D}_n$,
\[
\|g_n(\theta;D)-g_n(\theta;D')\|_2\le\Delta_n,
\qquad \Delta_n=\frac{2C_\phi M}{n}.
\]
\end{lemma}
The proof is in Appendix~\ref{app:score-sensitivity-proof}. Lemma \ref{lemma:sensitivity} guarantees that sampling $\theta$ from the KNG mechanism \eqref{eq:kng} is $\epsilon_\theta$-DP.

The next proposition shows that the weights evaluated at the privatized propensity score $e_{\tilde\theta}(X_i)$ retain quantitative control of covariate balance. For $z\in\{0,1\}$, we define
\[
 U_z(\theta)=\frac1n\sum_i1\{Z_i=z\}w_z\{e_\theta(X_i)\}\phi(X_i),\qquad
 \mathcal B_b(\theta)=\frac{U_1(\theta)}{\max\{B_1(\theta),b\}}
 -\frac{U_0(\theta)}{\max\{B_0(\theta),b\}},
\]
where $B_z$ is the weighted average of the arm indicator defined in Section~\ref{sec:cbsr}, and $b$ is the denominator floor, fixed independently of confidential records. The floor controls noise amplification in the ratio, since Laplace noise can move a denominator arbitrarily close to zero and make the ratio highly sensitive to small perturbations. Each denominator $B$, including after privatization, is replaced by $\max\{B,b\}$. The threshold is on the average scale of $B_z(\theta)$ and prevents division by small or negative noisy values. Larger $b$ limits noise amplification but can distort the estimate when binding. Under Assumption~\ref{asmp:identification}, a fixed choice $0<b<\eta m$, with $m$ from \eqref{eq:weight-bounds}, makes the unperturbed floors asymptotically inactive. Appendix~\ref{sec:finite_privacy} justifies this choice.

\begin{proposition}[Covariate balance]
\label{prop:finite_balance}
Under Assumption~\ref{asmp:public}, $U_1(\theta)-U_0(\theta)=-g_n(\theta)$ for every $\theta\in\Theta$. If the intercept belongs to the basis span, so $c^\top\phi(x)=1$ for a fixed vector $c$, then
\begin{equation}
 \|\mathcal B_b(\theta)\|_2
 \leq\frac{1+C_\phi\|c\|_2}{b}\|g_n(\theta)\|_2
 \quad(\theta\in\Theta).
 \label{eq:normalized_balance}
\end{equation}
Suppose additionally that $L_n$ is twice continuously differentiable on a neighborhood of $\Theta$, $g_n(\hat\theta)=0$, $B(\hat\theta,r_0)\subset\Theta$ for $r_0>0$, and
\[
 \kappa I_d\preceq\nabla^2L_n(\theta)\preceq\Lambda I_d
 \quad(\theta\in\Theta),\qquad 0<\kappa\leq\Lambda<\infty.
\]
With $a$ from \eqref{eq:kng}, let
\[
 F_d(t)=\tfrac1{\Gamma(d)}\int_0^t u^{d-1}\exp(-u)\,du,\;
 \chi_D=d\log\frac{2\Lambda}{\kappa}-\log F_d(a\Lambda r_0),\;
 K_D=\frac{2\sqrt{\chi_D^2+2\chi_D+2}}{a\kappa}.
\]
Then the privatized weights satisfy
\begin{equation}
 \E\{\|U_1(\tilde\theta)-U_0(\tilde\theta)\|_2^2\mid D\}
 \leq\Lambda^2K_D^2.
 \label{eq:balance_moment}
\end{equation}
If $a\Lambda r_0\geq2d$, $\chi_D$ can be replaced by $d\log(2\Lambda/\kappa)+\log2$.
\end{proposition}
Equation~\eqref{eq:normalized_balance} holds at the private draw $\tilde\theta$. Together with \eqref{eq:balance_moment}, whose expectation is conditional on the database, it controls average and arm-normalized imbalance. 
 These bounds quantify how private propensity fitting preserves the balancing property emphasized by \citet{Imai2014,Zhao2017,Jianqing2023}, even under propensity score model misspecification. 
 For fixed dimension, nondegenerate curvature, and an interior margin bounded away from zero, the average imbalance scales as $1/(n\epsilon_\theta)$. The proof is provided in Appendix~\ref{sec:finite_privacy}. 

\subsubsection{Private WATE Estimation}
We next consider the second step where we privatize the WATE estimator, using the privatized propensity score parameter $\tilde\theta$ obtained in the first step.
We define the arm mean and contrast as
\begin{equation}
\label{eq:protected-ratio}
R_b(A,B)=\clip_{[0,1]}\left\{\frac{A}{\max(B,b)}\right\},\qquad
T_b(\theta)=R_b(A_1(\theta),B_1(\theta))-R_b(A_0(\theta),B_0(\theta)).
\end{equation}
We release the four queries $Q=(A_1,B_1,A_0,B_0)$ at $\tilde\theta$, with independent noises
\begin{equation}
\label{eq:point-release}
\tilde Q_j=Q_j(\tilde\theta)+\nu_j,\quad
\nu_j\sim\Lap(0,\varsigma_j),\quad 
\varsigma_j=\frac{M}{n\epsilon_j},\quad
\tilde\tau_b=R_b(\tilde Q_1,\tilde Q_2)-R_b(\tilde Q_3,\tilde Q_4).
\end{equation}
For any fixed $\theta\in\Theta$, each query has sensitivity $M/n$, giving the Laplace scales in \eqref{eq:point-release}.
Clipping the protected arm means to $[0,1]$ keeps $\tilde\tau_b\in[-1,1]$ and guarantees finite mean squared error. Appendix~\ref{sec:finite_privacy} explains why unprotected noisy ratios can have infinite mean squared error. The following proposition provides a formal privacy guarantee of Algorithm \ref{alg:point}.

\begin{algorithm}[t]
\caption{Private covariate balancing with protected ratios}
\label{alg:point}
\begin{algorithmic}[1]
\REQUIRE Public $n,\phi,C_\phi,\Theta,\eta,\alpha,\beta,b,\epsilon_\theta,\epsilon_1,\ldots,\epsilon_4$; confidential records $D$.
\STATE Calculate $M$ from \eqref{eq:weight-bounds} and draw $\tilde\theta$ from \eqref{eq:kng}, using the exact sampler in Appendix~\ref{app:sampling}.
\STATE Internally compute $A_z(\tilde\theta)$ and $B_z(\tilde\theta)$ for $z=0,1$.
\STATE Draw the four noises and form $\tilde Q$ using \eqref{eq:point-release}.
\STATE Return $\tilde\tau_b$, and retain $\tilde\theta,\tilde Q$ if private inference is requested.
\end{algorithmic}
\end{algorithm}

\begin{proposition}[Privacy on the full record domain]
\label{prop:privacy}
Algorithm~\ref{alg:point}, including the joint release of $(\tilde\theta,\tilde Q)$, is $(\epsilon_\theta+\sum_{j=1}^4\epsilon_j)$-DP.
\end{proposition}
The proof is in Appendix~\ref{app:point-privacy-proof}.

\subsection{Theoretical Guarantees for the Private Point Estimator}
\label{sec:point-guarantees}\label{sec:privacy-cost}\label{sec:optimality}
In this section, we study conditional privacy cost, population risk, and the asymptotic distribution of the private point estimator. The limiting privacy cost also yields an oracle allocation of the point-estimation budget.

\subsubsection{Conditional Privacy Cost}
\label{sec:conditional-privacy}
We first compare the private estimator $\tilde\tau_b$ with the nonprivate estimator $\hat\tau_b=T_b(\hat\theta)$ conditional on the same database. The following theorem separates the perturbation from fitting the propensity score from that due to releasing the four weighted averages. 

\begin{theorem}[Conditional privacy cost]
\label{thm:conditional_wate}
Suppose the curvature and interior-root conditions of Proposition~\ref{prop:finite_balance} hold. The constants $a$, $\chi_D$, and $K_D$ are as in \eqref{eq:kng} and Proposition~\ref{prop:finite_balance}, and the Laplace scales are $\varsigma_j=M/(n\epsilon_j)$ from \eqref{eq:point-release}. Let
\[
\begin{gathered}
 G_D=\frac1n\sum_{i=1}^n\phi(X_i)\phi(X_i)^\top,\qquad
 L_D=2\min\left\{C_\phi,\sqrt{\frac{M\lambda_{\max}(G_D)}{b}}\right\},\\
 N_D=\frac{\sqrt{(\sum_{j=1}^4\varsigma_j)^2+\sum_{j=1}^4\varsigma_j^2}}{b}.
\end{gathered}
\]
Then
\begin{equation}
 \mathbb E\{(\tilde\tau_b-\hat\tau_b)^2\mid D\}
 \leq\min\{4,(L_DK_D+N_D)^2\}.
 \label{eq:conditional_privacy_cost}
\end{equation}
For $0<\delta<1$, with conditional probability at least $1-\delta$,
\begin{equation}
 |\tilde\tau_b-\hat\tau_b|
 \leq\min\left\{2,
 \frac{2L_D\{\chi_D+\log(2/\delta)\}}{a\kappa}
 +\frac{\log(8/\delta)}{b}\sum_{j=1}^4\varsigma_j\right\}.
 \label{eq:conditional_wate_tail}
\end{equation}
If $B_z(\theta)\geq\rho\geq b$ for both groups and every $\theta\in\Theta$, one may instead use
\[
L_D=\min\left\{C_\phi,\sqrt{\frac{M\lambda_{\max}(G_D)}{\rho}}\right\}.
\]
\end{theorem}
The proof is in Appendix~\ref{app:conditional-wate-proof}.
Furthermore, by Theorem~\ref{thm:conditional_kng} in Appendix \ref{sec:finite_privacy}, conditional on $D$, $K_D$ bounds the root mean squared Euclidean displacement of the KNG draw from the nonprivate fit, so $\mathbb E\{\|\tilde\theta-\hat\theta\|_2^2\mid D\}\leq K_D^2$. It measures privacy perturbation of the parameter, not its sampling error. The constant $L_D$ is a Lipschitz bound for the protected estimator, $|T_b(\theta)-T_b(\theta')|\leq L_D\|\theta-\theta'\|_2$ for all $\theta,\theta'\in\Theta$. Thus $L_DK_D$ translates parameter perturbation into a root mean squared perturbation of the weighted estimator, while $N_D$ accounts for the four Laplace releases. These conditional bounds do not represent an additive increase in population MSE. 
% Curvature and the interior radius are utility conditions. 

\subsubsection{Population Risk and Minimax Optimality}
Following the private minimax framework of \citet{barber2014}, we compare worst-case population MSE with a lower bound for all $\epsilon$-DP estimators on the same model class. As is standard in DP minimax analysis \citep{barber2014,Cai2021}, rate optimality means matching bounds jointly in $n$ and $\epsilon$, up to constants independent of both. Because these bounds hold uniformly in $n$ and $\epsilon$, they also apply when the public budget follows a predetermined sequence $\epsilon=\epsilon_n$ that decreases with sample size, providing stronger privacy as $n$ grows. 

The public objects in Assumption~\ref{asmp:public}, dimension $d$, estimand $(\alpha,\beta)\in[-1,0]^2$, and positive constants $g,r$ are fixed throughout the comparison. Suppose the closed ball $\overline B(0,r)$ lies in $\Theta$ and a distribution $P_X^\circ$ on the public covariate space satisfies
\[
 \mathbb E_{P_X^\circ}\{\phi(X)\phi(X)^\top\}\succeq gI_d.
\]
The class $\mathcal P_{\rm reg}$ consists of laws $P$ satisfying Assumption~\ref{asmp:identification}, with $Y(z)\in[0,1]$, for which
\begin{equation}
 \label{eq:regular-risk-class}
 e_0=e_{\theta_0},\qquad
 \mathbb E_P\{\phi(X)\phi(X)^\top\}\succeq gI_d,
 \qquad \overline B(\theta_0,r)\subseteq\Theta.
\end{equation}
The condition on $P_X^\circ$ ensures that the class is nonempty and supplies a full-rank design for the lower-bound submodel. The public denominator floor $b$ satisfies $0<b<\eta m$, with $m$ from \eqref{eq:weight-bounds}. The total point-estimation budget $\epsilon>0$ is divided into fixed public shares
\[
 \epsilon_\theta=\pi_0\epsilon,\qquad
 \epsilon_j=\pi_j\epsilon\ (1\le j\le4),\qquad
 \pi_k>0,\quad \sum_{k=0}^4\pi_k=1.
\]

For a matching comparison, let $\mathcal T_{n,\epsilon}$ contain all randomized real-valued estimators satisfying $\epsilon$-DP on that same record domain, and define the minimax risk for estimating $\tau_h(P)$ over $\mathcal P_{\rm reg}$ under squared error loss by
\[
 R^*_{n,\epsilon}(\mathcal P_{\rm reg})
 =\inf_{T\in\mathcal T_{n,\epsilon}}
   \sup_{P\in\mathcal P_{\rm reg}}
       \mathbb E_P\{(T-\tau_h(P))^2\}.
\]
A randomized estimator includes its privacy mechanism and subsequent processing.

The following minimax risk bound uses two explicit constants. The coefficient $C_{\rm stat}$ controls sampling error and the contribution of exceptional databases, while $C_{\rm priv}$ controls the contribution of KNG and Laplace perturbations. Their complete finite-sample formulas, including all auxiliary constants, are given in Appendix~\ref{app:risk-constants}, equations~\eqref{eq:risk-curvature-constants}--\eqref{eq:risk-upper-constants}.  Both constants are uniform over $\mathcal P_{\rm reg}$ and independent of $n$ and $\epsilon$. The appendix also retains a sharper bound with exponential remainder terms.

\begin{theorem}[Finite-sample upper and lower risk bounds]
\label{thm:uniform-risk}
For every integer $n\ge1$ and every $\epsilon>0$, the following bounds hold.

\textup{(i) Upper bound.} The protected estimator $\tilde\tau_b$ satisfies
\begin{equation}
\label{eq:uniform-risk}
 \sup_{P\in\mathcal P_{\rm reg}}
 \mathbb E_P\{(\tilde\tau_b-\tau_h(P))^2\}
 \le\min\left\{4,
 \frac{C_{\rm stat}}{n}+\frac{C_{\rm priv}}{n^2\epsilon^2}\right\}.
\end{equation}
The expectation includes sampling and mechanism randomness.

\textup{(ii) Lower bound.} The minimax risk over all estimators in $\mathcal T_{n,\epsilon}$ satisfies
\begin{equation}
\label{eq:lower-bound}
 R^*_{n,\epsilon}(\mathcal P_{\rm reg})
 \ge\max\left\{\frac1{256n},
       \frac3{32}\min\left(1,\frac1{n^2\epsilon^2}\right)\right\}.
\end{equation}
\end{theorem}

The proofs of parts (i) and (ii) are in Appendix~\ref{app:optimality-proofs}. The privacy guarantee holds for every database in the public record domain. Together, these bounds establish minimax-rate optimality jointly in $n$ and $\epsilon$, with common rate $n^{-1}+\min\{1,(n\epsilon)^{-2}\}$ and constants independent of both quantities. Appendix~\ref{app:minimax-comparison} gives the algebraic comparison. The following corollary states the explicit critical-budget comparison.

\begin{corollary}[Critical privacy budget]
\label{cor:joint-minimax}
For every $n\ge1$, the public deterministic budget $\epsilon_n=n^{-1/2}$ satisfies
\[
 \frac{3}{32n}
 \le R^*_{n,\epsilon_n}(\mathcal P_{\rm reg})
 \le\sup_{P\in\mathcal P_{\rm reg}}
       \mathbb E_P\{(\tilde\tau_b-\tau_h(P))^2\}
 \le\frac{C_{\rm stat}+C_{\rm priv}}{n}.
\]
Consequently, a privacy budget tending to zero at this rate retains the nonprivate mean squared error rate. For any public budget sequence satisfying $\sqrt n\epsilon_n\to0$, no sequence of $\epsilon_n$-DP estimators has a uniform $n^{-1}$ MSE rate over this class.
\end{corollary}

The proof is in Appendix~\ref{app:optimality-proofs}. At the critical budget, privacy can still contribute at first order to the limiting distribution. At fixed $\epsilon$, adding independent noise of variance $n^{-3/2}$ preserves total-risk rate optimality while increasing perturbation relative to an $n^{-2}$ privacy cost. The conditional discrepancy in \eqref{eq:conditional_privacy_cost} distinguishes these estimators. 
%Growing sieves are covered by the transfer result in Corollary~\ref{cor:privacy_transfer} in Section~\ref{sec:growing-series} under its dimension-dependent conditions and are outside this minimax theorem.

\subsubsection{Asymptotic Distribution and Oracle Budget Allocation}
\label{sec:joint_privacy}
The finite-sample bound \eqref{eq:conditional_privacy_cost} quantifies the perturbation of the estimator on a fixed database. We next give its leading constant and the allocation that minimizes this cost. These results also describe sampling uncertainty when the privacy budget decreases with sample size. Let
$L(\theta)=\E\{L_n(\theta)\}$.
The point-estimation budget has fixed public shares,
\begin{equation}
 \epsilon_{\theta,n}=\pi_0\epsilon_n,\qquad
 \epsilon_{j,n}=\pi_j\epsilon_n\quad(1\leq j\leq4),\qquad
 \pi_k>0,\quad\sum_{k=0}^4\pi_k=1.
 \label{eq:point_budget_shares}
\end{equation}
We allow $\sup_n\epsilon_n<\infty$ and $n\epsilon_n\to\infty$.

\begin{assumption}[Fixed-dimensional interior model]
\label{ass:fixed-joint}
The observations are iid and satisfy Assumptions~\ref{asmp:identification}
and \ref{asmp:public}, with fixed $d$, $\phi$, $\Theta$, $\eta$, and
$\alpha,\beta\in[-1,0]$. The true propensity equals $e_{\theta_0}$ for
$\theta_0\in\operatorname{int}(\Theta)$, and
$J=\nabla^2L(\theta_0)\succ0$. Let
$h_0(X)=e_{\theta_0}(X)^{\alpha+1}\{1-e_{\theta_0}(X)\}^{\beta+1}$,
$H=\E h_0(X)$, and $\mu_z=\E\{h_0(X)Y(z)\}/H$.
There is a $\delta>0$ such that $H\geq b+\delta$ and
$\delta\leq\mu_z\leq1-\delta$ for $z=0,1$.
\end{assumption}

Assumption~\ref{ass:fixed-joint} keeps the population denominator $H$ above the floor $b$ and the arm means $\mu_z$ away from $0$ and $1$. Flooring and clipping are therefore inactive near these population values, allowing a first-order expansion of the weighted contrast.

To describe the effect of perturbing the propensity parameter, define
\begin{equation}
\label{eq:joint_gamma}
\begin{aligned}
C_z
=\E\left[
1\{Z=z\}w_z\{e_{\theta_0}(X)\}
k_z\{e_{\theta_0}(X)\}(Y-\mu_z)\phi(X)
\right],\qquad
\gamma_\tau=(C_1-C_0)/H,
\end{aligned}
\end{equation}
where $k_z$ is the log-weight derivative in \eqref{eq:logweight-derivative}. 
Holding the data distribution fixed, $C_z/H$ is the gradient of the population weighted mean \eqref{eq:weighted_outcome_mean} with respect to the fitted propensity parameter, evaluated at $\theta_0$. Thus, $\gamma_\tau$ describes the first-order change in the weighted contrast due to a small parameter perturbation. 
Similarly, perturbations  $(\delta A_z,\delta B_z)$ to an arm's numerator and denominator change its weighted mean by $(\delta A_z-\mu_z\delta B_z)/H$ to first order.

These two derivatives determine how the privacy noise enters the limiting distribution. Let $W_d$ have density proportional to $\exp(-\|w\|_2)$ on $\mathbb R^d$, and let $L_1,\ldots,L_4$ be independent standard Laplace variables, independent of $W_d$. The limiting distribution of the scaled privacy perturbation $n\epsilon_n(\tilde\tau_b-\hat\tau_b)$ is represented by
\begin{equation}
\label{eq:joint_privacy_law}
L_\pi
=\frac{4C_\phi M}{\pi_0}\gamma_\tau^\top J^{-1}W_d
+\frac{M}{H}\left(
\frac{L_1}{\pi_1}
-\frac{\mu_1L_2}{\pi_2}
-\frac{L_3}{\pi_3}
+\frac{\mu_0L_4}{\pi_4}
\right).
\end{equation}
The first term captures the KNG propensity perturbation, with its effect determined by the loss curvature $J$ and the contrast derivative $\gamma_\tau$. The second term captures the four Laplace releases for $(A_1,B_1,A_0,B_0)$.
These components are independent and centered, so their variances add up. The following theorem identifies $\operatorname{Var}(L_\pi)$ as the leading constant in the conditional mean squared perturbation; minimizing this variance over the budget shares yields the oracle allocation.

\begin{theorem}[Joint sampling and privacy law]
\label{thm:joint-privacy-limit}
Under Assumption~\ref{ass:fixed-joint} and \eqref{eq:point_budget_shares}, the conditional law of $n\epsilon_n(\tilde\tau_b-\hat\tau_b)$ given $D$ converges weakly in probability to the law of $L_\pi$. Moreover,
\begin{equation}
 \left(\sqrt n(\hat\tau_b-\tau_h),\,
 n\epsilon_n(\tilde\tau_b-\hat\tau_b)\right)
 \xrightarrow{d}(G_V,L_\pi),\qquad G_V\sim N(0,V),
 \label{eq:joint_sampling_privacy_limit}
\end{equation}
where $V$ is the asymptotic variance coefficient of the nonprivate estimator $\hat\tau_b$, with sandwich expression given in \eqref{eq:sandwich}, and the limiting components are independent. The conditional squared
perturbation has the explicit limit
\begin{equation}
 n^2\epsilon_n^2\E\{(\tilde\tau_b-\hat\tau_b)^2\mid D\}
 \xrightarrow{p}C(\pi)=\sum_{k=0}^4\frac{a_k}{\pi_k^2}
 =\operatorname{Var}(L_\pi),
 \label{eq:joint_cost_limit}
\end{equation}
with
\begin{equation}
 a_0=16C_\phi^2M^2(d+1)\|J^{-\top}\gamma_\tau\|_2^2,
 \qquad (a_1,a_2,a_3,a_4)=\frac{2M^2}{H^2}(1,\mu_1^2,1,\mu_0^2).
 \label{eq:joint_cost_coefficients}
\end{equation}
\end{theorem}

The proof in Appendix~\ref{app:joint_privacy_proofs} extends the
fixed-dimensional KNG density argument of \citet[Theorem~3.2]{Reimherr2019}
to shrinking budgets and conditional moments through the protected ratios.
The cost in \eqref{eq:joint_cost_limit} measures perturbation relative to the
same estimator on the same database. 

In particular, if $\sqrt n\epsilon_n\to\infty$, the private estimator has
the same $N(0,V)$ root-$n$ limit as its nonprivate comparator. If
$\sqrt n\epsilon_n\to c_\epsilon\in(0,\infty)$, then
\[
 \sqrt n(\tilde\tau_b-\tau_h)\xrightarrow{d}G_V+c_\epsilon^{-1}L_\pi.
\]
Thus $\epsilon_n=n^{-1/2}$ preserves root-$n$ convergence while making
privacy increasingly stringent. If $\sqrt n\epsilon_n\to0$, then
$n\epsilon_n(\tilde\tau_b-\tau_h)\to_d L_\pi$.
 Corollary~\ref{cor:privacy_transfer} gives a separate
growing-basis transfer result when privacy perturbations are negligible at
the relevant scale.

\begin{corollary}[Oracle allocation of the point budget]
\label{cor:oracle_allocation}
Under Theorem~\ref{thm:joint-privacy-limit},
\begin{equation}
 \inf_{\pi_k>0,\,\sum_k\pi_k=1}C(\pi)
 =\left(\sum_{k=0}^4a_k^{1/3}\right)^3.
 \label{eq:oracle_cost}
\end{equation}
If every $a_k>0$, the unique minimizing shares are
$\pi_k^*=a_k^{1/3}/\sum_{m=0}^4a_m^{1/3}$.
If $a_0=0$, the infimum is approached as $\pi_0\downarrow0$ and is not
attained over strictly positive shares. With a public lower share
$0<\underline\pi<1/5$, the constrained optimum is
\[
 \pi_k^*=\max\{\underline\pi,(2a_k/\lambda)^{1/3}\},\qquad
 \sum_{k=0}^4\max\{\underline\pi,(2a_k/\lambda)^{1/3}\}=1,
\]
where the normalization uniquely determines $\lambda>0$.
\end{corollary}

The cube-root allocation minimizes the leading conditional mean squared privacy perturbation within the proposed mechanism. The earlier minimax result concerns worst-case estimation-error rates across all private estimators. Because the optimal shares depend on unknown population coefficients, they provide an oracle benchmark; Section~\ref{sec:simulation-allocation} discusses practical choices. When $\gamma_\tau=0$, propensity perturbation has no first-order effect on the weighted contrast, but higher-order effects may remain. The joint limit still relies on consistent private propensity fitting with a fixed positive budget share. Thus, the boundary infimum in Corollary~\ref{cor:oracle_allocation} does not justify setting the propensity budget to zero.

\subsection{Private Variance Estimation and Confidence Intervals}
\label{sec:variance_estimation}
\label{sec:inference}
We now develop differentially private variance estimation and confidence intervals for the proposed point estimator. We privately estimate the population quantities in Theorem~\ref{thm:joint-privacy-limit} and use the joint sampling and privacy law to construct intervals that account for both sampling variability and privacy noise.

We first characterize the sampling variability of the nonprivate estimator $\hat\tau_b$ through its influence function, whose variance is the coefficient $V$ in Theorem~\ref{thm:joint-privacy-limit}. Because the weighted arm means use propensity weights estimated from the same data, the influence function should account for their joint estimation. We obtain it by linearizing the joint estimating equations for the propensity parameter and the two weighted arm means. With $\xi=(\theta^\top,\mu_1,\mu_0)^\top$ and $e=e_\theta(X)$, define the estimating function
\begin{equation}
 \psi(O;\xi)=\begin{pmatrix}s(O;\theta)\\u_1(O;\xi)\\u_0(O;\xi)\end{pmatrix},
 \quad s=(Z-e)e^\alpha(1-e)^\beta\phi(X),\quad
 u_z=1\{Z=z\}w_z(e)(Y-\mu_z).
 \label{eq:stacked-score}
\end{equation}
Under Assumption~\ref{ass:fixed-joint}, the population parameter is $\xi_0=(\theta_0^\top,\mu_1,\mu_0)^\top$ and the sampling variance is
\begin{equation}
 \begin{gathered}
 V=v_\tau^\top J_\psi^{-1}\Sigma_\psi J_\psi^{-\top}v_\tau,
 \end{gathered}
 \label{eq:sandwich}
\end{equation}
where $J_\psi=\E\{\partial_\xi\psi(O;\xi_0)\}$ is the joint Jacobian, $\Sigma_\psi=\E\{\psi(O;\xi_0)\psi(O;\xi_0)^\top\}$ is the score second-moment matrix, and $v_\tau=(0_d^\top,1,-1)^\top$. 
Here $V/n$ is the first-order sampling variance. For the confidence-interval result below, we assume $V>0$, ensuring a nondegenerate sampling limit when privacy perturbations are asymptotically negligible. The subscript $\psi$ distinguishes the joint Jacobian $J_\psi$ from the propensity-loss Hessian $J$. The corresponding influence function is
\begin{equation}
 \operatorname{IF}_\tau(O)=(u_1-u_0)/H+\gamma_\tau^\top J^{-1}s,
 \label{eq:wate-influence-function}
\end{equation}
with all quantities evaluated at $\xi_0$.
The second term accounts for propensity estimation, including its effect
on a propensity-dependent target population. Appendix~\ref{app:inference-proofs} gives the
matrix derivatives and the influence-function calculation.

To estimate the sandwich matrices privately, let
$\tilde\xi=(\tilde\theta^\top,\tilde\mu_1,\tilde\mu_0)^\top$,
using the private propensity estimate and protected arm means from
the point-estimation procedure. For a public feature bound
$C_{\phi,1}\geq\sup_x\|\phi(x)\|_1$ and additional public privacy
budgets $\epsilon_{\mathrm{jac}},\epsilon_{\mathrm{mom}}>0$, release
\begin{equation}
\begin{aligned}
\tilde J_\psi
&=\frac1n\sum_i\partial_\xi\psi(O_i;\tilde\xi)
  +E_{\mathrm{jac}},
&
(E_{\mathrm{jac}})_{jk}
&\sim\Lap\left(
0,\frac{2M(C_{\phi,1}^2+C_{\phi,1}+1)}
        {n\epsilon_{\mathrm{jac}}}
\right),\\
\tilde\Sigma_\psi
&=\frac1n\sum_i
  \psi(O_i;\tilde\xi)\psi(O_i;\tilde\xi)^\top
  +E_{\mathrm{mom}},
&
(E_{\mathrm{mom}})_{jk}
&\sim\Lap\left(
0,\frac{2M^2(C_{\phi,1}+1)^2}
        {n\epsilon_{\mathrm{mom}}}
\right).
\end{aligned}
\label{eq:private-sandwich-moments}
\end{equation}
All noise entries are independent. The scales use
$\ell_1$-sensitivity bounds for the entire vectorized matrix averages,
with $\tilde\xi$ held fixed.
Appendix~\ref{app:inference-proofs} gives the derivation.

The added noise can destroy the symmetry and positive semidefiniteness of the second-moment matrix and make Jacobian inversion unstable. We therefore symmetrize $\tilde\Sigma_\psi$ and replace its negative eigenvalues by zero, obtaining $\tilde\Sigma_{\psi,+}$. To stabilize matrix inversion, we use a positive threshold $\zeta_n$, chosen independently of the confidential data and satisfying $\zeta_n\to0$. Our implementation uses $\zeta_n=n^{-1/4}$. The regularized inverse $\tilde J_{\psi,\zeta}^{-1}$ is obtained by replacing each singular value $\sigma$ of $\tilde J_\psi$ with $\max\{\sigma,\zeta_n\}$ before inversion. This prevents division by very small singular values that can arise after adding privacy noise. These adjustments give the well-defined, nonnegative private estimate of $V$,
\begin{equation}
\tilde V
=v_\tau^\top
 \tilde J_{\psi,\zeta}^{-1}
 \tilde\Sigma_{\psi,+}
 \tilde J_{\psi,\zeta}^{-\top}
 v_\tau.
\label{eq:private-sandwich}
\end{equation}

The same releases also provide estimates of the coefficients
governing the privacy perturbations in
\eqref{eq:joint_privacy_law}. Obtain $\tilde J$ by symmetrizing
the negative propensity block
$-(\tilde J_\psi)_{1:d,1:d}$ and replacing eigenvalues below
$\zeta_n$ by $\zeta_n$. The remaining estimates are
\begin{equation}
\begin{gathered}
\tilde C_1
=\{(\tilde J_\psi)_{d+1,1:d}\}^\top,\qquad
\tilde C_0
=\{(\tilde J_\psi)_{d+2,1:d}\}^\top,\\
\tilde H_1=\max(\tilde Q_2,b),\qquad
\tilde H_0=\max(\tilde Q_4,b),\qquad
\tilde\gamma_\tau
=\tilde C_1/\tilde H_1-\tilde C_0/\tilde H_0.
\end{gathered}
\label{eq:private-noise-coefficients}
\end{equation}

Substituting these estimates into the joint sampling and privacy
law gives a reference distribution for
$\tilde\tau_b-\tau_h$ that combines sampling variability with the
KNG and Laplace perturbations from point estimation.
Independently of the private releases, generate mutually
independent variables $Z_{\rm ref}\sim N(0,1)$,
$W_d$ as in \eqref{eq:joint_privacy_law}, and standard Laplace
variables $L_1,\ldots,L_4$, and define
\begin{equation}
\begin{split}
\widehat U_n={}
\sqrt{\tilde V/n}\,Z_{\rm ref}
+\frac{4C_\phi M}{n\epsilon_{\theta,n}}
 \tilde\gamma_\tau^\top\tilde J^{-1}W_d+\frac{M}{n\tilde H_1}
 \left(
 \frac{L_1}{\epsilon_{1,n}}
 -\frac{\tilde\mu_1L_2}{\epsilon_{2,n}}
 \right)
-\frac{M}{n\tilde H_0}
 \left(
 \frac{L_3}{\epsilon_{3,n}}
 -\frac{\tilde\mu_0L_4}{\epsilon_{4,n}}
 \right).
\end{split}
\label{eq:full-noise-reference}
\end{equation}

For $0<\alpha_{\rm CI}<1$, let $q_{1-\alpha_{\rm CI},n}$ be the
conditional $(1-\alpha_{\rm CI})$ quantile of $|\widehat U_n|$
given the private releases. In practice, this quantile can be
approximated by the empirical quantile of independent simulated
values of $|\widehat U_n|$, holding the released coefficients
fixed. The confidence interval is
\begin{equation}
I_n=
[\tilde\tau_b-q_{1-\alpha_{\rm CI},n},\,
 \tilde\tau_b+q_{1-\alpha_{\rm CI},n}]
\cap[-1,1].
\label{eq:private-confidence-interval}
\end{equation}
The matrix adjustments, coefficient calculations, and reference
simulation are post-processing of the private releases and
require no additional privacy budget. The following theorem
establishes differential privacy of the complete procedure,
consistency of the coefficient estimates, and pointwise
asymptotic coverage.
\begin{theorem}[Private inference]
\label{thm:private-inference}
Releasing $(\tilde\theta,\tilde Q,\tilde J_\psi,\tilde\Sigma_\psi,I_n)$ is
$(\epsilon_n+\epsilon_{\mathrm{jac}}+\epsilon_{\mathrm{mom}})$-DP on the public record domain.
Under Assumption~\ref{ass:fixed-joint} and the fixed positive point-budget
shares in \eqref{eq:point_budget_shares}, suppose
$n\epsilon_n,n\epsilon_{\mathrm{jac}},n\epsilon_{\mathrm{mom}}\to\infty$ and
$\sup_n\epsilon_n<\infty$. Then
\[
 (\tilde V,\tilde J,\tilde\gamma_\tau,
  \tilde H_1,\tilde H_0,\tilde\mu_1,\tilde\mu_0)
 \xrightarrow{p}(V,J,\gamma_\tau,H,H,\mu_1,\mu_0).
\]
For fixed $0<\alpha_{\rm CI}<1$,
$\Pr\{\tau_h\in I_n\}\to1-\alpha_{\rm CI}$ in each regime
$\sqrt n\epsilon_n\to\infty$,
$\sqrt n\epsilon_n\to c_\epsilon\in(0,\infty)$, or
$\sqrt n\epsilon_n\to0$. 
% The coverage conclusion also holds for the empirical reference quantile from $N_{{\rm ref},n}\to\infty$ independent reference draws.
\end{theorem}

If $\sqrt n\epsilon_n\to\infty$, privacy noise is asymptotically negligible, and the interval reduces to the sandwich Wald interval. When $\epsilon_n\asymp n^{-1/2}$, the KNG and Laplace perturbations are of the same order as sampling error whenever their coefficients are nonzero, and therefore affect the interval width.  Appendix~\ref{app:inference-proofs} provides the proof, a bound controlling simulation error in the interval quantile, and a budget allocation under which the fraction spent on the two matrix releases tends to zero.

% End section formerly in inference.tex

\subsection{Extension to Growing Series}
\label{sec:growing-series}

Finally, we allow the number of basis functions to increase with sample size. A richer basis expands the collection of covariate functions whose imbalance is controlled and may improve approximation of the propensity score. We give sufficient conditions under which the private estimator retains the consistency and limiting distribution of its nonprivate counterpart.

Let $\phi_n$ have dimension $d_n$ and parameter domain $\Theta_n$. The basis, parameter domain, denominator floor, and privacy budgets must be public choices fixed independently of confidential observations. The finite-sample bound in Theorem~\ref{thm:conditional_wate} continues to apply when these choices vary with $n$. In particular, $L_DK_D+N_D$ bounds the conditional root mean squared difference between the private and nonprivate estimators. Consistency is preserved when this perturbation vanishes; a root-$n$ limiting distribution is preserved when it vanishes faster than $n^{-1/2}$.

\begin{corollary}[Transfer of consistency and sampling limits]
\label{cor:privacy_transfer}
Allow the public objects and constants in
Theorem~\ref{thm:conditional_wate} to vary with $n$.
Suppose its conditions hold on sample events $\mathcal E_n$
with $\Pr(\mathcal E_n)\to1$.

\begin{enumerate}
    \item (Consistency) If $L_DK_D+N_D=o_p(1)$ on these events and $\hat\tau_b\xrightarrow{p}\tau_{\rm target}$, then $\tilde\tau_b\xrightarrow{p}\tau_{\rm target}$.
    \item (Sampling limit) If $L_DK_D+N_D=o_p(n^{-1/2})$ on these events and $\sqrt n(\hat\tau_b-\tau_{\rm target}) \xrightarrow{d} N(0,V),$ then $\sqrt n(\tilde\tau_b-\tau_{\rm target}) \xrightarrow{d}N(0,V)$.
\end{enumerate}
\end{corollary}

The proof is in Appendix~\ref{app:transfer-proof}. To obtain explicit conditions on basis growth, let $C_{\phi,n}\geq\sup_x\|\phi_n(x)\|_2$ be a public bound on the size of the basis vector. Suppose that $M$, $b^{-1}$, $\kappa^{-1}$, $\Lambda$, and $\lambda_{\max}(G_D)$ are bounded by constants independent of $n$ on $\mathcal E_n$, and that $a\Lambda r_0\geq2d_n$. On these events, the finite-sample bound gives
\[
 L_DK_D+N_D
 \leq C\left\{
 \frac{C_{\phi,n}d_n}{n\epsilon_{\theta,n}}
 +\sum_{j=1}^4\frac{1}{n\epsilon_{j,n}}
 \right\},
\]
for a constant $C$ independent of $n$. The first term bounds the effect of private propensity fitting, and the second bounds the noise from the four weighted-summary releases. Thus, sufficient conditions for preserving a nonprivate root-$n$ limit are
\begin{equation}
 \frac{C_{\phi,n}d_n}{\sqrt n\epsilon_{\theta,n}}
 \longrightarrow0,
 \qquad
 \frac{1}{\sqrt n\epsilon_{j,n}}
 \longrightarrow0
 \quad(j=1,\ldots,4).
 \label{eq:sieve_privacy_condition}
\end{equation}

For fixed dimension and bounded $C_{\phi,n}$, these conditions allow every privacy budget to tend to zero, provided $\sqrt n\epsilon_{\theta,n}\to\infty$ and $\sqrt n\epsilon_{j,n}\to\infty$. Preserving the same limiting distribution requires privacy noise to be negligible relative to sampling error. This is stronger than retaining the $n^{-1}$ mean squared error rate at the critical budget $\epsilon_n=n^{-1/2}$ in Corollary~\ref{cor:joint-minimax}, where privacy noise can still contribute at first order.

When the basis grows, both its dimension and the size of its basis vectors matter. For a basis orthonormal under a probability reference measure, the average squared norm of $\phi_n(x)$ is $d_n$, so $C_{\phi,n}\geq\sqrt{d_n}$. Two illustrative cases give the following requirements for the propensity-fitting budget:
\[
\begin{aligned}
 C_{\phi,n}\asymp\sqrt{d_n}
 :
 \frac{d_n^{3/2}}{\sqrt n\epsilon_{\theta,n}}\longrightarrow0,\qquad
 C_{\phi,n}\asymp d_n
 :
 \frac{d_n^2}{\sqrt n\epsilon_{\theta,n}}\longrightarrow0.
\end{aligned}
\]
These examples still require the curvature and interior-root conditions of the corollary; orthonormality alone does not guarantee them.

Applying the corollary therefore combines these privacy conditions with separately established approximation and sampling results for the chosen nonprivate series estimator \citep{hirano2003,Zhao2019}. The preceding minimax, oracle allocation, and private inference results retain their fixed-dimensional assumptions.

% Begin section formerly in empirical.tex
% Replacement empirical evidence. All results were generated and verified for v1.1.

\section{Simulation Studies}
\label{sec:simulation}
\label{sec:simulations}
\subsection{Setup}
\label{sec:simulationsetting}
We compare private CBSR with nonprivate CBSR and standard logistic IPW, private KNG-logistic, and the methods of \citet{Lee2019} and \citet{guha2024} under correctly specified and misspecified propensity models, evaluating bias, RMSE, covariate imbalance, and the coverage and mean width of nominal 95\% confidence intervals. We additionally examine private CBSR under correct specification at $n\in\{10^4,10^5\}$ and $\epsilon_{\rm total}\in\{1,5\}$, reporting bias, MSE, coverage, and mean interval width. 

All methods use the same covariate basis, public parameter domain, and protected weighted summaries. We set the overlap bound to $\eta=0.2$, the weight bound to $M=5$, and the denominator floor to $b=0.1$. The parameter domain is the Euclidean ball centered at zero with radius $\log(4)/C_\phi$. In these experiments, 80\% of $\epsilon_{\rm total}$ is divided equally among propensity fitting and the four weighted-summary releases. The remaining 20\% is divided equally between the two matrix releases used for variance estimation. Appendix~\ref{sec:simulation_details} provides full details of the simulation design, including the data-generating models, comparator implementations, and computational diagnostics.

\subsection{Correctly Specified Propensity Model}
We use four independent covariates uniform on $\{-1,0,1\}$. The fitted basis contains these covariates and an intercept. Table~\ref{tab:simulation_well_specified} reports the private CBSR results for two sample sizes and two total privacy budgets, using 1,000 independent replications per configuration. Table~\ref{tab:simulation_well_specified} shows that MSE and interval width decrease with both sample size and privacy budget. At $n=10^4$, increasing $\epsilon_{\rm total}$ from 1 to 5 reduces MSE from $3.02\times10^{-4}$ to $8.72\times10^{-5}$ and mean interval width from 1.3215 to 0.0549, while coverage decreases from 1.000 to 0.978. At $n=10^5$, MSE is $1.06\times10^{-5}$ and $8.53\times10^{-6}$ at budgets 1 and 5, respectively. Coverage is close to the nominal 0.95 at both budgets, at 0.956 and 0.952, with mean interval widths 0.0146 and 0.0115. Additional results and implementation details appear in Appendix~\ref{sec:simulation_details}.

\begin{table}[htbp]
\centering\small
\caption{Evaluation metrics of the proposed method under correctly specified propensity score model.}
\label{tab:simulation_well_specified}
\label{tab:correct-budget-diagnostics}
\begin{tabular}{rrrrrr}
\toprule
$n$ & $\epsilon_{\rm total}$ & Bias & MSE & Coverage & Mean width\\
\midrule
10,000 & 1 & 0.003640 & $3.02\times10^{-4}$ & 1.000 & 1.3215\\
10,000 & 5 & -0.000159 & $8.72\times10^{-5}$ & 0.978 & 0.0549\\
100,000 & 1 & 0.000061 & $1.06\times10^{-5}$ & 0.956 & 0.0146\\
100,000 & 5 & 0.000210 & $8.53\times10^{-6}$ & 0.952 & 0.0115\\
\bottomrule
\end{tabular}
\end{table}

\subsection{Misspecified Propensity Model}
We consider two scenarios in which the fitted logistic propensity model omits a higher-order term in the true treatment log odds. In the linear outcome scenario, the fitted basis $(1,X)^\top$ excludes a quadratic propensity term. In the quadratic outcome scenario, the fitted basis $(1,X,X^2)^\top$ excludes a cubic propensity term. The conditional outcome means lie in the respective basis spans in both scenarios, allowing us to assess the benefit of covariate balance when the propensity model is misspecified. Table~\ref{tab:simulation_misspecified} also compares \citet{Lee2019} and \citet{guha2024}. The private methods use a budget of 1.6 for point estimation and an additional 0.4 for interval construction where available. \citet{Lee2019} provides point estimates only under $(1.6,10^{-6})$-DP, while the other complete private procedures satisfy pure $2$-DP.

Table~\ref{tab:simulation_misspecified} shows that private CBSR has the smallest RMSE among the displayed private methods in both scenarios, at 0.0029 for the linear outcome and 0.0031 for the quadratic outcome. Its imbalance is also smaller than that of KNG-logistic, at 0.0012 versus 0.0261 in the linear scenario and 0.0027 versus 0.0053 in the quadratic scenario. Private CBSR coverage is 0.957 and 0.942, compared with 0.328 and 0.919 for KNG-logistic. Guha--Reiter coverage is 0.955 and 0.942, with mean interval lengths 0.0757 and 0.0758, compared with 0.0117 and 0.0121 for private CBSR. Additional results and implementation details appear in Appendix~\ref{app:empirical-comparators}.

\begin{table}[htbp]
\centering\small
\caption{Propensity misspecification with $n=10^5$, using 1,000 linear and 2,000 quadratic datasets paired across methods. NP is nonprivate. DP budgets are 1.6 for points and 0.4 for intervals. Lee is $(1.6,10^{-6})$-DP without intervals. Other complete releases are pure 2-DP. Intervals have nominal 95\% coverage and are intersected with $[-1,1]$. Balance is the norm of arm-normalized feature-mean differences. $\dagger$ denotes the Guha--Reiter interval adaptation in Appendix~\ref{app:empirical-comparators}.}
\label{tab:simulation_misspecified}
\setlength{\tabcolsep}{5pt}
\begin{tabular}{lrrrrr}
\toprule
Estimator & Bias & RMSE & Balance & Coverage & Mean length\\
\midrule
\multicolumn{6}{l}{Linear outcome}\\
CBSR, NP & -0.0001 & 0.0029 & 0.0000 & 0.955 & 0.0115\\
\textbf{CBSR, DP (Ours)} & -0.0001 & 0.0029 & 0.0012 & 0.957 & 0.0117\\
Logistic, NP & -0.0071 & 0.0077 & 0.0261 & 0.322 & 0.0116\\
KNG-logistic, DP & -0.0071 & 0.0077 & 0.0261 & 0.328 & 0.0117\\
Lee et al., DP & 0.0015 & 0.0063 & -- & -- & --\\
Guha--Reiter$^\dagger$, DP & -0.0068 & 0.0189 & -- & 0.955 & 0.0757\\
\midrule
\multicolumn{6}{l}{Quadratic outcome}\\
CBSR, NP & 0.0001 & 0.0030 & 0.0000 & 0.945 & 0.0116\\
\textbf{CBSR, DP (Ours)} & 0.0001 & 0.0031 & 0.0027 & 0.942 & 0.0121\\
Logistic, NP & -0.0013 & 0.0033 & 0.0051 & 0.917 & 0.0116\\
KNG-logistic, DP & -0.0013 & 0.0034 & 0.0053 & 0.919 & 0.0119\\
Lee et al., DP & -0.0540 & 0.0544 & -- & -- & --\\
Guha--Reiter$^\dagger$, DP & -0.0013 & 0.0184 & -- & 0.942 & 0.0758\\
\bottomrule
\end{tabular}
\end{table}

\subsection{Oracle Budget Allocation}
\label{sec:simulation-allocation}
Theorem~\ref{thm:joint-privacy-limit} identifies $C(\pi)=\sum_{k=0}^4a_k/\pi_k^2$ as the leading scaled privacy-perturbation cost. Corollary~\ref{cor:oracle_allocation} proves that the cube-root shares minimize this cost when the population coefficients are known. These claims concern conditional privacy perturbation within our mechanism, rather than exact finite-sample total MSE or interval length.

We illustrate this allocation in the two-dimensional correctly specified model described in Appendix~\ref{app:empirical-designs}. The point budget is $\epsilon_{\rm point}=0.8$. At each sample size, 20 independent databases receive 150 independent mechanism draws for each allocation. Applying the cube-root formula to the known generating distribution gives shares for $(\theta,A_1,B_1,A_0,B_0)$ of
\[
 (0.35935,0.21030,0.12166,0.21030,0.09839).
\]
The resulting leading scaled cost is 5,376, compared with 9,106 under five equal shares, a predicted reduction of 41\%. These costs approximate $n^2\epsilon_{\rm point}^2\E\{(\tilde\tau_b-\hat\tau_b)^2\mid D\}$. Thus, at $n=10^4$ and $\epsilon_{\rm point}=0.8$, the oracle predicts squared privacy perturbation $5376/(10^4\times0.8)^2=8.4\times10^{-5}$.

Table~\ref{tab:simulation-allocation} compares the predicted and empirical scaled costs. Oracle allocation reduces the empirical cost by 36.9\%--39.4\% across the three sample sizes, consistent with the predicted 41.0\% reduction. Table~\ref{tab:r3-finite-bound} in the appendix gives unscaled costs, Monte Carlo standard errors, and the explicit finite-sample bounds.

\begin{table}[htbp]
\centering\small
\caption{Scaled privacy-perturbation cost under equal and oracle budget allocations, with $\epsilon_{\rm point}=0.8$. The asymptotic prediction is $C(\pi)$. Empirical costs estimate $n^2\epsilon_{\rm point}^2\E\{(\tilde\tau_b-\hat\tau_b)^2\mid D\}$, averaged over 20 independent databases with 150 mechanism draws per database and allocation. Parentheses give Monte Carlo standard errors across databases. Reduction is relative to equal allocation.}
\label{tab:simulation-allocation}
\begin{tabular}{lrrr}
\toprule
Setting & Equal allocation & Oracle allocation & Reduction\\
\midrule
Asymptotic & 9,106 & 5,376 & 41.0\%\\
\midrule
Empirical, $n=10^4$ & 8,838 (240) & 5,355 (172) & 39.4\%\\
Empirical, $n=10^6$ & 8,726 (318) & 5,367 (157) & 38.5\%\\
Empirical, $n=10^8$ & 8,728 (240) & 5,503 (176) & 36.9\%\\
\bottomrule
\end{tabular}
\end{table}

When the population coefficients are unknown, equal shares provide a simple default. Cube-root shares can instead be estimated from suitable independent public data or a private pilot, with a positive minimum share for each release to stabilize the allocation. A private pilot can estimate the coefficients using the releases in Section~\ref{sec:inference}. Computing shares from these private outputs requires no additional privacy budget, but the pilot's privacy cost must be included in the total budget. The final estimator then uses the remaining budget and independent randomness. The oracle allocation provides a benchmark for these estimated allocations.

\section{Data Analysis}
\label{sec:analysis}
\label{sec:application}
We apply the method to the email campaign data of \citet{Hillstrom2008Public}. The treatment is assignment to receive an email featuring women's merchandise ($Z=1$), with no email as the control condition ($Z=0$). The binary outcome $Y$ equals one if the customer visited the retailer's website during the two weeks following the campaign and zero otherwise. There were 3,238 website visitors in the treatment group and 2,262 in the control group. The intended estimand is the average effect of the email on the probability of a website visit among customers eligible for the campaign, under consistency and no interference. One customer is the privacy unit. This application examines how privacy protection affects covariate balance and covariate-adjusted point and interval estimates, using the nonprivate CBSR fit as a benchmark.

Adjustment uses an intercept, purchase recency, annual spending, and prior purchase of women's merchandise. Appendix~\ref{app:customer-details} specifies the public covariate transformations and analysis bounds. At each total privacy budget $\epsilon_{\rm total}\in\{1,5,10\}$, 80\% is allocated to point estimation and 20\% to inference. Each of the five point-estimation releases receives $0.16\epsilon_{\rm total}$, and each of the two matrix releases receives $0.10\epsilon_{\rm total}$.

Table~\ref{tab:customer-private} reports the estimate, confidence interval, and interval width from the same first scheduled draw at each privacy budget. The private estimates range from 4.512 to 4.634 percentage points, close to the nonprivate estimate of 4.519, and all displayed intervals exclude zero. The interval width is 3.465 at budget 1, compared with 1.261 nonprivately. At budgets 5 and 10, widths of 1.300 and 1.393 are closer to the nonprivate width. Widths from individual draws need not decrease monotonically with budget because their privately estimated coefficients are random. RMSE over all 100 draws decreases from 0.141 to 0.029 and 0.017 percentage points as the budget increases.

\begin{table}[htbp]
\centering\small
\caption{Email campaign effects. Estimates, confidence limits, widths, and RMSE are in percentage points. At each budget, the estimate, 95\% confidence interval, and its width use the same first scheduled draw. RMSE uses all 100 independent privatizations and compares with the nonprivate CBSR estimate. All 300 CBSR draws completed.}
\label{tab:ate_ci}
\label{tab:customer-private}
\begin{tabular}{lrrrr}
\toprule
$\epsilon_{\rm total}$ & Estimate & 95\% confidence interval & Interval width & RMSE\\
\midrule
Nonprivate & 4.519 & $[3.888, 5.149]$ & 1.261 & -- \\
1 & 4.634 & $[2.902, 6.367]$ & 3.465 & 0.141 \\
5 & 4.571 & $[3.921, 5.221]$ & 1.300 & 0.029 \\
10 & 4.512 & $[3.816, 5.209]$ & 1.393 & 0.017 \\
\bottomrule
\end{tabular}
\end{table}

Table~\ref{tab:customer-balance} reports $\|\mathcal B_b(\tilde\theta)\|_2$, the Euclidean norm of the difference between the two arms' weighted feature means, using the weights from the private propensity draw. The three covariates are on their publicly specified $[-1,1]$ scales, and the intercept contributes zero because the denominator floors are inactive. The nonprivate fit balances these features to numerical precision. Mean imbalance over 100 private draws decreases from 0.02197 at budget 1 to 0.00407 at budget 5 and 0.00222 at budget 10, with corresponding 90th percentiles of 0.03650, 0.00709, and 0.00361. These diagnostics evaluate the private weights on public data. Publishing the same diagnostics from confidential records would require additional privacy protection. Additional results and implementation details appear in Appendix~\ref{app:customer-details}.

\begin{table}[htbp]
\centering\small
\caption{Covariate balance in the email campaign. Imbalance is $\|\mathcal B_b(\tilde\theta)\|_2$ on the transformed feature scale, with smaller values indicating better balance. The first-draw values use the same private propensity draws as Table~\ref{tab:customer-private}. Means and 90th percentiles use all 100 draws at each budget. The nonprivate row reports the single fitted value.}
\label{tab:customer-balance}
\begin{tabular}{lrrr}
\toprule
$\epsilon_{\rm total}$ & First-draw imbalance & Mean imbalance & 90th percentile\\
\midrule
Nonprivate & $1.40\times10^{-14}$ & -- & -- \\
1 & 0.00936 & 0.02197 & 0.03650 \\
5 & 0.00237 & 0.00407 & 0.00709 \\
10 & 0.00080 & 0.00222 & 0.00361 \\
\bottomrule
\end{tabular}
\end{table}

\section{Concluding Remarks}
\label{sec:conclusion}
This paper develops a differentially private approach to covariate balancing for weighted average treatment effect estimation. The two-stage procedure protects both propensity fitting and the weighted outcome and denominator queries. Explicit finite-sample bounds quantify the resulting covariate imbalance and estimation error. For regular, correctly specified propensity models of fixed dimension, matching risk bounds establish minimax-rate optimality jointly in sample size and privacy budget. The joint sampling and privacy law identifies an oracle budget allocation within the proposed mechanism and supports inference using privately estimated moments. Simulations and the email campaign application illustrate how privacy protection affects covariate balance and the recovery of nonprivate point and interval estimates.

Future research should broaden robustness and improve the practical implementation of private covariate balancing. Although the score-based balance identity applies under propensity misspecification, balancing a finite set of features alone does not ensure consistent effect estimation. Private feature selection and richer balancing bases merit study with explicit control of approximation error, sensitivity, and curvature. Another direction is inference with reliable finite-sample coverage when overlap is weak or parameter constraints bind, extending the pointwise asymptotic guarantees established here. Practical budget allocation also warrants investigation when the oracle coefficients are unknown, with the privacy cost of estimating these coefficients included in the analysis. Finally, computationally efficient samplers with certified privacy guarantees and private balance diagnostics could support repeated model assessment while accounting for its cumulative privacy cost.

\clearpage
\appendix
\begin{center}\Large\bfseries Supplementary material\end{center}

\section{Supporting perturbation bounds and proofs}
\label{sec:finite_privacy}
This appendix gives the KNG perturbation and prognostic-error bounds and proves the privacy, conditional-cost, and growing-series transfer results. We retain the public objects in Assumption~\ref{asmp:public} and the KNG coefficient $a=n\epsilon_\theta/(4C_\phi M)$ from \eqref{eq:kng}. Conditional probabilities and expectations concern mechanism randomness given the database.

\begin{theorem}[Conditional KNG perturbation]
\label{thm:conditional_kng}
Suppose that, for a fixed database, $L_n$ is twice continuously differentiable on a neighborhood of $\Theta$ and
\[
 \kappa I_d\preceq\nabla^2L_n(\theta)\preceq\Lambda I_d
 \quad(\theta\in\Theta),\qquad 0<\kappa\leq\Lambda<\infty.
\]
Suppose $g_n(\hat\theta)=0$ and the Euclidean ball $B(\hat\theta,r_0)$ is contained in $\Theta$ for some $r_0>0$. Let
\[
 F_d(t)=\frac{1}{\Gamma(d)}\int_0^t u^{d-1}\exp(-u)\,du,
 \quad \chi_D=d\log\frac{2\Lambda}{\kappa}-\log F_d(a\Lambda r_0).
\]
For $R=\|\tilde\theta-\hat\theta\|_2$ and every $0<\delta<1$,
\begin{align}
 \Pr\left\{R>\frac{2\{\chi_D+\log(1/\delta)\}}{a\kappa}\,\middle|\,D\right\}&\leq\delta,
 \label{eq:kng_tail}\\
 \mathbb E(R^2\mid D)&\leq K_D^2,
 \qquad K_D=\frac{2\sqrt{\chi_D^2+2\chi_D+2}}{a\kappa}.
 \label{eq:kng_moment}
\end{align}
If $a\Lambda r_0\geq2d$, then $F_d(a\Lambda r_0)\geq1/2$, and $\chi_D$ can be replaced in these bounds by $d\log(2\Lambda/\kappa)+\log2$.
\end{theorem}

The proof is in Appendix~\ref{app:finite-proofs}.
The curvature conditions in Theorem~\ref{thm:conditional_kng} are conditions on the realized database. They are needed for utility, whereas the privacy guarantee holds on every admissible database, including databases without an interior root or a nonsingular empirical design. The constants in a conditional bound need not be released. In particular, estimating them from confidential data does not authorize their publication or their use to select an unaccounted privacy budget.

For the Beta family, the weights are $w_1(e)=e^\alpha(1-e)^{\beta+1}$ and $w_0(e)=e^{\alpha+1}(1-e)^\beta$. Direct differentiation gives
\begin{align}
 c_1(e)&=w_1(e)\{(\beta+1)e-\alpha(1-e)\},\nonumber\\
 c_0(e)&=w_0(e)\{(\alpha+1)(1-e)-\beta e\},\nonumber\\
 \nabla^2L_n(\theta)&=\frac1n\sum_{i=1}^n c_{Z_i}\{e_\theta(X_i)\}\phi(X_i)\phi(X_i)^\top.
 \label{eq:cbsr_curvature}
\end{align}
Thus empirical Gram eigenvalues, rather than population orthogonality alone, determine curvature. For ATE, the coefficients lie between $\eta/(1-\eta)$ and $(1-\eta)/\eta$. For ATO they equal $e(1-e)$, which lies between $\eta(1-\eta)$ and $1/4$. For ATT only control observations contribute to the Hessian, and for ATC only treated observations contribute. These two estimands can therefore satisfy the theorem when the corresponding group Gram matrix is positive definite. They should not be excluded solely because one of their individual score contributions is linear.

\begin{proposition}[Prognostic-error extension]
\label{prop:prognostic_error}
Suppose the curvature and interior-root conditions of Proposition~\ref{prop:finite_balance} hold and the intercept belongs to the basis span, with $c^\top\phi(x)=1$. Assume $B_z(\theta)\geq\rho\geq b$ for both groups and all $\theta\in\Theta$, and
\[
 m_1(x)-m_0(x)=\tau_0,\qquad
 \sup_x|m_0(x)-\gamma^\top\phi(x)|\leq r_{\rm app},
 \quad m_z(x)=\mathbb E\{Y(z)\mid X=x\}.
\]
Under unconfoundedness, the conditional mean of the first-stage weighted contrast obeys
\begin{equation}
 \left|\mathbb E\{T_b(\tilde\theta)\mid X_{1:n},Z_{1:n},\tilde\theta\}-\tau_0\right|
 \leq\frac{\|\gamma\|_2+(1+r_{\rm app})\|c\|_2}{\rho}
 \|g_n(\tilde\theta)\|_2+2r_{\rm app}.
 \label{eq:prognostic_bias}
\end{equation}
\end{proposition}

The proof is in Appendix~\ref{app:finite-proofs}.

Equation~\eqref{eq:prognostic_bias} describes the effect of privacy on adjustment for an approximable baseline response. It does not establish consistency under arbitrary propensity misspecification and arbitrary treatment-effect heterogeneity. Under heterogeneity, the same argument controls baseline imbalance relative to the treated group's weighted average of $m_1(X_i)-m_0(X_i)$, which is generally distinct from the population WATE. 

The already privatized denominators can sharpen the second-stage part of \eqref{eq:conditional_wate_tail} without further access to the database. On the event $|\nu_j|\leq t_j=\varsigma_j\log(8/\delta)$, a sharper second-stage bound is
\[
 \min\left\{1,\frac{t_1+t_2}{\max\{\tilde Q_2,b\}}\right\}
 +\min\left\{1,\frac{t_3+t_4}{\max\{\tilde Q_4,b\}}\right\}.
\]
The perturbation calculation in the proof retains each protected noisy denominator before bounding it below by $b$. In contrast, using confidential empirical curvature constants to publish a first-stage radius requires separate privacy protection or valid public bounds.

Stabilization is necessary for a squared-error analysis. For the unmodified estimator, a Laplace-perturbed denominator has positive density at zero. An independent numerator has positive second moment, so the squared ratio has infinite expectation. Convergence in probability of the raw ratio cannot justify a finite mean squared error or a mean squared error rate.

For the asymptotic choice $0<b<\eta m$, let $n_z=\sum_i1\{Z_i=z\}$. The weight lower bound gives $B_z(\theta)\geq m n_z/n$ for every $\theta\in\Theta$. Under Assumption~\ref{asmp:identification}, $n_z/n\to P(Z=z)\geq\eta$ almost surely, so both unperturbed denominators exceed $b$ uniformly over $\Theta$ with probability tending to one. The finite-sample bounds retain $b$ explicitly. Both the private estimator and its nonprivate comparator use the same floor. Any error from applying this floor before privatization is included in the nonprivate comparator's error.

For a finite-sample population-risk comparison at any target $\tau_{\rm target}$, if $\mathcal E_n$ has probability at least $1-\delta_n^{\rm bad}$ and $L_DK_D+N_D\leq q_n$ there, then
\[
 \mathbb E\{(\tilde\tau_b-\hat\tau_b)^2\}\leq q_n^2+4\delta_n^{\rm bad},
 \qquad
 \{\mathbb E(\tilde\tau_b-\tau_{\rm target})^2\}^{1/2}
 \leq\{\mathbb E(\hat\tau_b-\tau_{\rm target})^2\}^{1/2}
 +\sqrt{q_n^2+4\delta_n^{\rm bad}}.
\]
This statement retains a separately defined privacy perturbation term and does not identify it with an additive increase in population mean squared error.

\subsection{Proofs of Privacy and Conditional Bounds}
\label{app:finite-proofs}\label{app:privacy-proofs}

\subsubsection{Proof of Lemma~\ref{lemma:sensitivity}}
\label{app:score-sensitivity-proof}
\begin{proof}
Each row's score vector has norm at most $C_\phi M$. Replacing a row changes the average score vector by at most twice this bound divided by $n$. The supremum covers both possible records and is not a function of a realized $X_i$ or $Z_i$.
\end{proof}

\subsubsection{Proof of Proposition~\ref{prop:privacy}}
\label{app:point-privacy-proof}
\begin{proof}
The reverse triangle inequality and Lemma~\ref{lemma:sensitivity} bound the change in the exponent of \eqref{eq:kng} by $a\Delta_n=\epsilon_\theta/2$. Integrating the corresponding inequality bounds the ratio of normalizing constants by $\exp(\epsilon_\theta/2)$. Hence the likelihood ratio of the first-stage distributions is at most $\exp(\epsilon_\theta)$. For every fixed first-stage output $\tilde\theta$, each $Q_j(\tilde\theta)$ is an average of terms in $[0,M]$. Replacing one record therefore changes the query by at most $M/n$. The scales in \eqref{eq:point-release} give privacy cost $\sum_j\epsilon_j$ for the four Laplace releases. Adaptive composition then applies even though they use the same records. The final protected ratios are post-processing.
\end{proof}
\subsubsection{Proof of Theorem~\ref{thm:conditional_kng}}
\begin{proof}
Strong convexity, smoothness, and $g_n(\hat\theta)=0$ imply
\[
 \kappa\|\theta-\hat\theta\|_2\leq\|g_n(\theta)\|_2
 \leq\Lambda\|\theta-\hat\theta\|_2.
\]
Let $s_{d-1}$ be the surface area of the unit sphere. The normalizing integral of the KNG density is at least
\[
 s_{d-1}\int_0^{r_0}r^{d-1}\exp(-a\Lambda r)\,dr
 =\frac{s_{d-1}\Gamma(d)}{(a\Lambda)^d}F_d(a\Lambda r_0).
\]
For the proof, write $\omega=a\kappa/2$. The integral defining the numerator of $\mathbb E\{\exp(\omega R)\mid D\}$ is bounded above by integration over all of $\mathbb R^d$, giving
\[
 \mathbb E\{\exp(\omega R)\mid D\}
 \leq\frac{(2\Lambda/\kappa)^d}{F_d(a\Lambda r_0)}=\exp(\chi_D).
\]
Markov's inequality proves \eqref{eq:kng_tail}. Integrating the bound $\Pr(R>r\mid D)\leq\min\{1,\exp(\chi_D-\omega r)\}$ gives
\[
 \mathbb E(R^2\mid D)
 \leq\int_0^{\chi_D/\omega}2r\,dr+
 \int_{\chi_D/\omega}^{\infty}2r \exp(\chi_D-\omega r)\,dr
 =\frac{\chi_D^2+2\chi_D+2}{\omega^2}.
\]
Finally, a $\operatorname{Gamma}(d,1)$ random variable has mean $d$, so Markov's inequality gives $F_d(2d)\geq1/2$.
\end{proof}

\subsubsection{Proof of Proposition~\ref{prop:finite_balance}}
\begin{proof}
The score identity follows directly from \eqref{eq:gradient}. Since $g_n(\hat\theta)=0$, the Hessian upper bound gives $\|g_n(\tilde\theta)\|_2\leq\Lambda R$, where $R=\|\tilde\theta-\hat\theta\|_2$. The conditional second-moment bound in Theorem~\ref{thm:conditional_kng} proves \eqref{eq:balance_moment}, including its stated simplification. Write $d_z=\max\{B_z,b\}$. Then
\[
 \mathcal B_b=\frac{U_1-U_0}{d_1}
 +\frac{U_0}{d_0}\frac{d_0-d_1}{d_1}.
\]
Use $\|U_0/d_0\|_2\leq C_\phi$, $|d_1-d_0|\leq|B_1-B_0|$, and $B_1-B_0=c^\top(U_1-U_0)$ to obtain \eqref{eq:normalized_balance}.
\end{proof}

\subsubsection{Proof of Proposition~\ref{prop:prognostic_error}}
\begin{proof}
For \eqref{eq:prognostic_bias}, conditioning on the private propensity parameter does not change the conditional outcome means because KNG uses only covariates and treatment assignments. The constant effect contributes $\tau_0$, and the two approximation remainders contribute at most $2r_{\rm app}$. The remaining term is $\gamma^\top\mathcal B_b$. With inactive floors it equals
\[
 \frac{\gamma^\top(U_1-U_0)
 -(\gamma^\top U_0/B_0)c^\top(U_1-U_0)}{B_1}.
\]
Since $0\leq m_0\leq1$, the approximation condition gives $|\gamma^\top U_0/B_0|\leq1+r_{\rm app}$. The asserted bound follows.
\end{proof}

\subsubsection{Proof of Theorem~\ref{thm:conditional_wate}}
\label{app:conditional-wate-proof}
\begin{proof}
Let $t=\theta^\top\phi(x)$. The two logarithmic weight derivatives satisfy
\[
 \frac{d\log w_1}{dt}=\alpha(1-e)-(\beta+1)e\in[-1,0],\qquad
 \frac{d\log w_0}{dt}=(\alpha+1)(1-e)-\beta e\in[0,1].
\]
Where $B_z>b$, the gradient of the group ratio is a weighted average of
\[
 \{Y_i-A_z(\theta)/B_z(\theta)\}\nabla\log w_z.
\]
Because $Y_i\in[0,1]$, its norm is at most $C_\phi/2$. Cauchy--Schwarz and the bound on the weighted variance of $Y_i$ also give the upper bound $\tfrac12\{M\lambda_{\max}(G_D)/B_z\}^{1/2}$. Where $B_z<b$, the ratio is $A_z/b$ and its gradient has norm at most $C_\phi A_z/b\leq C_\phi$. A second Cauchy--Schwarz bound gives
\[
 \|\nabla A_z\|_2^2\leq B_z M\lambda_{\max}(G_D),
\]
so the ratio gradient is at most $\{M\lambda_{\max}(G_D)/b\}^{1/2}$. The map is continuous and Lipschitz along segments, including where $B_z=b$. Summing the two group bounds proves that $T_b(\theta)$ is $L_D$-Lipschitz. If the floors are inactive throughout $\Theta$, the sharper bound follows.

For $0\leq A\leq B$, nonexpansiveness of projection and of $u\mapsto\max(u,b)$ gives
\[
 |R_b(A+v_A,B+v_B)-R_b(A,B)|
 \leq\frac{|v_A|+|v_B|}{b}.
\]
Indeed, writing $d=\max(B,b)$ and $d'=\max(B+v_B,b)$, one has $A/d\leq1$ and the difference before projection is bounded by $(|v_A|+(A/d)|d-d'|)/d'$. Therefore
\[
 |\tilde\tau_b-\hat\tau_b|\leq L_D R+\frac1b\sum_{j=1}^4|\nu_j|.
\]
The absolute value of a centered Laplace variable of scale $\varsigma_j$ has mean $\varsigma_j$ and variance $\varsigma_j^2$. Independence and Minkowski's inequality establish \eqref{eq:conditional_privacy_cost}. Apply \eqref{eq:kng_tail} with error probability $\delta/2$ and the four Laplace tail bounds $\Pr\{|\nu_j|>\varsigma_j\log(8/\delta)\}=\delta/8$ to obtain \eqref{eq:conditional_wate_tail}. Both estimators lie in $[-1,1]$, proving the additional deterministic bounds.
\end{proof}

\subsubsection{Proof of Corollary~\ref{cor:privacy_transfer}}
\label{app:transfer-proof}
\begin{proof}
Conditional Markov's inequality applied to \eqref{eq:conditional_privacy_cost}, followed by averaging over the data, gives $\tilde\tau_b-\hat\tau_b=o_p(1)$ in the first case and $o_p(n^{-1/2})$ in the second. More explicitly, truncate the conditional probability bound at one and use convergence in probability together with boundedness to take its expectation. The complements of $\mathcal E_n$ have vanishing probability. Slutsky's theorem completes the transfer argument.

For the sufficient conditions in \eqref{eq:sieve_privacy_condition}, the bounded constants and $a\Lambda r_0\geq2d_n$ give $\chi_D\leq d_n\log(2\Lambda/\kappa)+\log2$, so $K_D$ is bounded by a constant times $C_{\phi,n}d_n/(n\epsilon_{\theta,n})$. The Gram bound keeps $L_D$ bounded independently of dimension, and $N_D$ is bounded by a constant times $\sum_j(n\epsilon_{j,n})^{-1}$. Finally, orthonormality under a probability reference measure $P_{\rm ref}$ implies $d_n=\int\|\phi_n(x)\|_2^2\,dP_{\rm ref}(x)\leq C_{\phi,n}^2$. This proves the envelope claim used in the discussion.
\end{proof}

\section{Proofs of uniform risk and minimax optimality}
\label{app:optimality-proofs}
\subsection{Uniform Population Risk}
\label{app:risk-constants}
The constants in Theorem~\ref{thm:uniform-risk} are defined explicitly below. For the Hessian coefficients
\[
 c_1(e)=w_1(e)\{(\beta+1)e-\alpha(1-e)\},\qquad
 c_0(e)=w_0(e)\{(\alpha+1)(1-e)-\beta e\},
\]
the uniform curvature and concentration constants are
\begin{equation}
\label{eq:risk-curvature-constants}
\begin{gathered}
 \kappa=\frac{\eta g}{2}\left\{
       \min_{e\in[\eta,1-\eta]}c_1(e)+\min_{e\in[\eta,1-\eta]}c_0(e)\right\}>0,
 \qquad \Lambda=MC_\phi^2,\\
 a_G=\frac{\kappa^2}{2d^2M^2C_\phi^4},\qquad
 a_S=\frac{\kappa^2r^2}{8dM^2C_\phi^2},\\
 \chi_*=d\log(2\Lambda/\kappa)+\log2,\qquad
 t_*=\frac{16dC_\phi M}{\pi_0\Lambda r}.
\end{gathered}
\end{equation}
The two risk coefficients are
\begin{equation}
\label{eq:risk-upper-constants}
\begin{aligned}
 C_{\rm stat}
 &=16M^2\left(b^{-2}+C_\phi^4/\kappa^2\right)
       +\frac{8d^2}{\exp(1)a_G}+\frac{8d}{\exp(1)a_S},\\
 C_{\rm priv}
 &=\max\left[\frac{1024C_\phi^4M^2(\chi_*^2+2\chi_*+2)}{\pi_0^2\kappa^2}
 +\frac{4M^2}{b^2}\left\{
       \left(\sum_{j=1}^4\pi_j^{-1}\right)^2
       +\sum_{j=1}^4\pi_j^{-2}\right\},\ 4t_*^2\right].
\end{aligned}
\end{equation}
The first two terms inside the maximum account for the KNG and query perturbations. The term $4t_*^2$ extends the bound to every positive privacy budget. The last two terms in $C_{\rm stat}$ absorb exponentially small probabilities of exceptional databases. All constants are uniform over $\mathcal P_{\rm reg}$ and independent of $n$ and $\epsilon$.

\begin{proof}[Proof of Theorem~\ref{thm:uniform-risk}(i)]
Write $g_n=\nabla L_n$, where $L_n$ is the average negative CBSR score, and let $\hat\theta$ be a measurable constrained minimizer on $\Theta$. Compactness guarantees existence on every database. Ties can be resolved by any fixed measurable rule.

First, the logarithmic weight derivatives are
\[
 k_1(e)=\alpha(1-e)-(\beta+1)e\in[-1,0],\qquad
 k_0(e)=(\alpha+1)(1-e)-\beta e\in[0,1].
\]
Consequently, the Hessian coefficients satisfy $0\le c_z(e)\le M$. Each coefficient is strictly positive on $[\eta,1-\eta]$ unless it vanishes identically. The latter case occurs for $c_1$ only at $(\alpha,\beta)=(0,-1)$ and for $c_0$ only at $(-1,0)$. They cannot both vanish, so $\kappa>0$.

Put $\underline c_z=\min_{e\in[\eta,1-\eta]}c_z(e)$ and define the weighted empirical Gram matrix
\[
 S_n=\frac1n\sum_{i=1}^n\underline c_{Z_i}\phi(X_i)\phi(X_i)^\top.
\]
True overlap and \eqref{eq:regular-risk-class} give
\[
 \mathbb E_P S_n
 =\mathbb E_P[\{e_0(X)\underline c_1+(1-e_0(X))\underline c_0\}
                  \phi(X)\phi(X)^\top]
 \succeq2\kappa I_d.
\]
Each entry of a summand lies in $[-MC_\phi^2,MC_\phi^2]$. Hoeffding's inequality and a union bound over the $d^2$ entries imply
\begin{equation}
\label{eq:risk-gram-event}
 \Pr_P\{\|S_n-\mathbb E_P S_n\|_{\rm op}>\kappa\}
 \le2d^2\exp(-a_Gn),
\end{equation}
using $\|A\|_{\rm op}\le d\max_{j,k}|A_{jk}|$. On the complementary event $\mathcal E_G$, for every $\theta\in\Theta$,
\begin{equation}
\label{eq:risk-empirical-curvature}
 \kappa I_d\preceq S_n\preceq\nabla^2L_n(\theta)
 \preceq\Lambda I_d.
\end{equation}
The lower comparison follows term by term from $c_{Z_i}\{e_\theta(X_i)\}\ge\underline c_{Z_i}$, and the upper bound uses $\|\phi(X_i)\|_2\le C_\phi$.

Correct propensity specification gives $\mathbb E_P g_n(\theta_0)=0$, because
\[
 e_0 w_1(e_0)=(1-e_0)w_0(e_0)=h_0.
\]
Each individual score vector has norm at most $MC_\phi$, hence
\begin{equation}
\label{eq:risk-score-moment}
 \mathbb E_P\|g_n(\theta_0)\|_2^2\le\frac{M^2C_\phi^2}{n}.
\end{equation}
Coordinatewise Hoeffding bounds also give
\begin{equation}
\label{eq:risk-score-event}
 \Pr_P\{\|g_n(\theta_0)\|_2>\kappa r/2\}
 \le2d\exp(-a_Sn).
\end{equation}
Let $\mathcal E$ be the intersection of $\mathcal E_G$ and the complementary score event in \eqref{eq:risk-score-event}. Then
\begin{equation}
\label{eq:risk-good-probability}
 \Pr_P(\mathcal E^c)\le2d^2\exp(-a_Gn)+2d\exp(-a_Sn).
\end{equation}

The following argument establishes interiority without assuming a score root. On $\mathcal E_G$, constrained optimality gives
$g_n(\hat\theta)^\top(\hat\theta-\theta_0)\le0$. Strong monotonicity in \eqref{eq:risk-empirical-curvature} therefore yields
\[
 \kappa\|\hat\theta-\theta_0\|_2^2
 \le\{g_n(\hat\theta)-g_n(\theta_0)\}^\top(\hat\theta-\theta_0)
 \le\|g_n(\theta_0)\|_2\|\hat\theta-\theta_0\|_2.
\]
Thus
\begin{equation}
\label{eq:risk-fit-bound}
 \|\hat\theta-\theta_0\|_2\le\|g_n(\theta_0)\|_2/\kappa.
\end{equation}
On $\mathcal E$, this distance is at most $r/2$. It follows that $\overline B(\hat\theta,r/2)\subseteq\Theta$, the minimizer is interior, and $g_n(\hat\theta)=0$.

Write $t=n\epsilon$ and first assume $t\ge t_*$. For the KNG density, $a=n\epsilon_\theta/(4C_\phi M)$, so on $\mathcal E$ its interior-ball constant obeys
\[
 a\Lambda(r/2)=\frac{t\pi_0\Lambda r}{8C_\phi M}\ge2d.
\]
Theorem~\ref{thm:conditional_kng}, with $r_0=r/2$, then gives
\begin{equation}
\label{eq:risk-kng-moment}
 \mathbb E\{\|\tilde\theta-\hat\theta\|_2^2\mid D\}
 \le\frac{64C_\phi^2M^2(\chi_*^2+2\chi_*+2)}
              {t^2\pi_0^2\kappa^2}
 \qquad(D\in\mathcal E).
\end{equation}

We next bound the four parts of the estimation error. At the true propensity, let $A_z=A_z(\theta_0)$ and $B_z=B_z(\theta_0)$ denote empirical averages, and set
\[
 H=\mathbb E_Ph_0(X),\qquad
 \mu_z=\mathbb E_P\{h_0(X)Y(z)\}/H.
\]
The means of $A_z,B_z$ are $H\mu_z,H$. Each summand is in $[0,M]$, so each empirical average has variance at most $M^2/(4n)$. Also $H\ge\eta m>b$. The protected ratio map satisfies
\[
 |R_b(A',B')-R_b(A,B)|
 \le\frac{|A'-A|+|B'-B|}{b}
 \quad\text{when }0\le A\le B,
\]
as established in the proof of Theorem~\ref{thm:conditional_wate}. Comparing each empirical pair with $(H\mu_z,H)$ and applying the inequality $(\sum_{j=1}^4|x_j|)^2\le4\sum_{j=1}^4x_j^2$ gives
\begin{equation}
\label{eq:risk-oracle-sampling}
 \mathbb E_P\{(T_b(\theta_0)-\tau_h)^2\}
 \le\frac{4M^2}{b^2n}.
\end{equation}

The Lipschitz calculation in Appendix~\ref{app:conditional-wate-proof} gives $L_D\leq2C_\phi$, including where a denominator floor binds. Equations~\eqref{eq:risk-score-moment} and \eqref{eq:risk-fit-bound} imply
\begin{equation}
\label{eq:risk-fitted-sampling}
 \mathbb E_P[\{T_b(\hat\theta)-T_b(\theta_0)\}^2 1_{\mathcal E}]
 \le\frac{4M^2C_\phi^4}{\kappa^2n}.
\end{equation}
Similarly, \eqref{eq:risk-kng-moment} gives
\begin{equation}
\label{eq:risk-first-privacy}
 \mathbb E_P[\{T_b(\tilde\theta)-T_b(\hat\theta)\}^2 1_{\mathcal E}]
 \le\frac{256C_\phi^4M^2(\chi_*^2+2\chi_*+2)}
              {t^2\pi_0^2\kappa^2}.
\end{equation}

Finally, the four independent Laplace noises have scales $\varsigma_j=M/(t\pi_j)$. The protected ratio bound gives
\[
 |\tilde\tau_b-T_b(\tilde\theta)|\le b^{-1}\sum_{j=1}^4|\nu_j|.
\]
Since $|\nu_j|$ has mean $\varsigma_j$ and variance $\varsigma_j^2$,
\begin{equation}
\label{eq:risk-query-privacy}
 \mathbb E_P\{(\tilde\tau_b-T_b(\tilde\theta))^2\}
 \le\frac{M^2}{b^2t^2}
   \left\{\left(\sum_{j=1}^4\pi_j^{-1}\right)^2
                   +\sum_{j=1}^4\pi_j^{-2}\right\}.
\end{equation}
The bound holds on every database.

Decompose $\tilde\tau_b-\tau_h$ into the four differences in \eqref{eq:risk-oracle-sampling}--\eqref{eq:risk-query-privacy}. On $\mathcal E$, its square is at most four times the sum of their squares. On $\mathcal E^c$, the squared error is at most $4$, because both $\tilde\tau_b$ and $\tau_h$ belong to $[-1,1]$. Combining the preceding bounds yields
\begin{equation}
\label{eq:refined-uniform-risk}
\begin{split}
 \mathbb E_P\{(\tilde\tau_b-\tau_h)^2\}
 &\le\min\biggl\{4,\frac{16M^2(b^{-2}+C_\phi^4/\kappa^2)}{n}
       +\frac{C_{\rm priv}}{t^2}\\
 &\qquad+8d^2\exp(-a_Gn)+8d\exp(-a_Sn)\biggr\}.
\end{split}
\end{equation}
For $t\ge t_*$, $C_{\rm priv}$ in this bound may be replaced by the first argument of the maximum in \eqref{eq:risk-upper-constants}. For $0<t<t_*$, the deterministic bound $4$ is at most $4t_*^2/t^2\leq C_{\rm priv}/t^2$, so \eqref{eq:refined-uniform-risk} holds for every $t>0$. Finally, $\exp(-an)\leq1/\{\exp(1)an\}$ for every $a>0$ and $n\geq1$. Absorbing the two exponential terms into $C_{\rm stat}/n$, taking the supremum over the common class, and intersecting with the deterministic bound proves \eqref{eq:uniform-risk}.
\end{proof}

\subsection{A Matching Lower Bound on the Same Class}
\begin{proof}[Proof of Theorem~\ref{thm:uniform-risk}(ii)]
Fix a distribution $P_X^\circ$ allowed in the definition of $\mathcal P_{\rm reg}$. Consider $X\sim P_X^\circ$, $Z\sim\operatorname{Bernoulli}(1/2)$, $Y(0)=0$, and $Y(1)\sim\operatorname{Bernoulli}(q)$, with $X,Z,Y(1)$ independent. The true propensity has parameter $\theta_0=0$. The design Gram matrix is at least $gI_d$ and $\overline B(0,r)\subseteq\Theta$, so this entire submodel lies in $\mathcal P_{\rm reg}$. Its tilt is constant and $\tau_h=q$ for all $\alpha,\beta\in[-1,0]$. The observed $(Z,Y)$ probabilities are $1/2,q/2,(1-q)/2$ on $(0,0),(1,1),(1,0)$, with the same independent covariate law at every $q$.

For parameters $q_0,q_1$ separated by $u=|q_1-q_0|$ and output distributions $\mathsf P_0,\mathsf P_1$, testing which parameter is closer to the estimate gives
\begin{equation}
\label{eq:regular-two-point}
 \max_{j=0,1}\mathbb E_j(T-q_j)^2
 \ge\frac{u^2}{8}\{1-\operatorname{TV}(\mathsf P_0,\mathsf P_1)\}.
\end{equation}
For the sampling term, choose $q_0=1/2$, $q_1=1/2+1/(4\sqrt n)$. The one-record Kullback--Leibler divergence is one half of the Bernoulli divergence. Using $\operatorname{KL}\{\operatorname{Bernoulli}(1/2+u),\operatorname{Bernoulli}(1/2)\}\le4u^2$, the divergence of the $n$ observations is at most $2nu^2=1/8$. Pinsker's inequality and data processing give output total variation at most $1/4$, hence at most $1/2$. Equation~\eqref{eq:regular-two-point} gives $1/(256n)$ for any randomized estimator.

For the privacy term, take $q_0=0$ and $q_1=u=\min\{1,(n\epsilon)^{-1}\}$. The one-record total variation is $u/2$. Couple the covariates identically and couple the remaining record components maximally and independently across observations. The Hamming distance of the coupled databases is $N_{\mathrm{diff}}\sim\operatorname{Binomial}(n,u/2)$. For databases at distance $N_{\mathrm{diff}}$, group privacy bounds likelihood ratios of their output distributions in both directions by $\exp(\epsilon N_{\mathrm{diff}})$. Two laws with this likelihood-ratio bound have total variation at most $\tanh(\epsilon N_{\mathrm{diff}}/2)\le\epsilon N_{\mathrm{diff}}/2$. Convexity of total variation under mixing and the coupling therefore give
\[
 \operatorname{TV}(\mathsf P_0,\mathsf P_1)
 \le\epsilon\mathbb E N_{\mathrm{diff}}/2
 =n\epsilon u/4\le1/4.
\]
Applying \eqref{eq:regular-two-point} yields $(3/32)\min\{1,(n\epsilon)^{-2}\}$. The supremum risk on the class exceeds each submodel risk. Taking the larger bound and then the infimum over all private estimators proves the result.
\end{proof}

\subsection{Minimax Comparison and the Critical Privacy Budget}
\label{app:minimax-comparison}
\begin{proof}[Joint minimax comparison and proof of Corollary~\ref{cor:joint-minimax}]
Let $r_{n,\epsilon}=n^{-1}+\min\{1,(n\epsilon)^{-2}\}$. The joint comparison is, for every $n\ge1$ and $\epsilon>0$,
\begin{equation}
\label{eq:joint-minimax-rate}
 \frac1{512}r_{n,\epsilon}
 \le R^*_{n,\epsilon}(\mathcal P_{\rm reg})
 \le\sup_{P\in\mathcal P_{\rm reg}}
        \mathbb E_P\{(\tilde\tau_b-\tau_h(P))^2\}
 \le\max\{4,C_{\rm stat},C_{\rm priv}\}\,r_{n,\epsilon}.
\end{equation}
For $n\epsilon\ge1$, the upper comparison follows directly from Theorem~\ref{thm:uniform-risk}(i). For $n\epsilon<1$, it follows from the deterministic bound $4$. For nonnegative $x,y$, $\max\{x/256,3y/32\}\ge(x+y)/512$. Theorem~\ref{thm:uniform-risk}(ii) with $x=n^{-1}$ and $y=\min\{1,(n\epsilon)^{-2}\}$ proves the lower comparison in \eqref{eq:joint-minimax-rate}. The middle comparison follows because the proposed estimator is itself $\epsilon$-DP on the full public record domain.

At $\epsilon_n=n^{-1/2}$, substituting into Theorem~\ref{thm:uniform-risk}(i) gives the upper bound $(C_{\rm stat}+C_{\rm priv})/n$, and the privacy part of the explicit lower bound equals $3/(32n)$. This proves the displayed threshold bounds. If instead $\sqrt n\epsilon_n\to0$, then
\[
 n\min\{1,(n\epsilon_n)^{-2}\}
 =\min\{n,(n\epsilon_n^2)^{-1}\}\longrightarrow\infty.
\]
Thus no sequence of mechanisms satisfying those budgets can have a uniform $n^{-1}$ mean squared error bound over $\mathcal P_{\rm reg}$.
\end{proof}

% End section formerly in supplement/optimality_proof.tex
% Begin section formerly in supplement/joint_limit_proofs.tex

\section{Proof of the joint privacy law and allocation}
\label{app:joint_privacy_proofs}
\begin{lemma}[Conditional density and moment convergence]
\label{lem:kng-density}
Under the assumptions of Theorem~\ref{thm:joint-privacy-limit}, let $p_n$
be the conditional Lebesgue density of
$n\epsilon_{\theta,n}(\tilde\theta-\hat\theta)$ given $D$, and let $p$
be the density of $4C_\phi M J^{-1}W_d$. For every fixed $q\geq0$,
\begin{equation}
 \int_{\mathbb R^d}(1+\|v\|_2^q)|p_n(v)-p(v)|\,dv\xrightarrow{p}0.
 \label{eq:kng_weighted_density_limit}
\end{equation}
\end{lemma}

\begin{proof}[Proof of Lemma~\ref{lem:kng-density} and Theorem~\ref{thm:joint-privacy-limit}]
We first verify the empirical regularity conditions, then prove conditional convergence on typical data sequences. All norms below are Euclidean or induced operator norms unless specified otherwise.

\paragraph{Population curvature and empirical roots.}
The Hessian identity \eqref{eq:cbsr_curvature} has nonnegative scalar coefficients $c_1(e)=-w_1(e)k_1(e)$ and $c_0(e)=w_0(e)k_0(e)$. For $\alpha,\beta\in[-1,0]$, $c_1$ vanishes identically only at $(\alpha,\beta)=(0,-1)$, and $c_0$ only at $(-1,0)$. They cannot vanish simultaneously at any $e\in[\eta,1-\eta]$. Hence
\[
 c_*:=\min_{\eta\leq e\leq1-\eta}\{c_1(e)+c_0(e)\}>0.
\]
Conditioning on $X$ gives, for every $\theta\in\Theta$,
\begin{equation}
 \nabla^2 L(\theta)\succeq\eta c_*\,\mathbb E\{\phi(X)\phi(X)^\top\}.
 \label{eq:joint_global_curvature}
\end{equation}
At $\theta_0$, conditioning the score derivative gives
\[
 J=\mathbb E\{h_0(X)\phi(X)\phi(X)^\top\}.
\]
Since $h_0$ is bounded and $J\succ0$, $\mathbb E(\phi\phi^\top)\succ0$. Thus \eqref{eq:joint_global_curvature} has a fixed positive lower eigenvalue. Also $\nabla L(\theta_0)=0$ by correct specification. Global strong convexity makes $\theta_0$ the unique population minimizer. For ATT and ATC, the conditional treatment probabilities in \eqref{eq:joint_global_curvature} retain the relevant arm contribution.

Bounded $\phi$ and overlap imply that the logistic predictor and all the derivatives used here are uniformly bounded on a fixed neighborhood of $\Theta$. The loss, its Hessian, the four queries and their first derivatives have bounded Lipschitz envelopes. The strong law on successively finer finite nets of $\Theta$, followed by the uniform Lipschitz bound, proves their uniform strong laws. The uniform loss law and uniqueness imply $\hat\theta\to\theta_0$ almost surely. The uniform Hessian law and \eqref{eq:joint_global_curvature} imply that, almost surely for all sufficiently large $n$, there are deterministic constants $\kappa,\Lambda,r>0$ such that
\begin{equation}
 \kappa I_d\preceq\nabla^2L_n(\theta)\preceq\Lambda I_d
 \ (\theta\in\Theta),\quad
 B(\hat\theta,r)\subset\Theta,\quad g_n(\hat\theta)=0,
 \quad \nabla^2L_n(\hat\theta)\longrightarrow J.
 \label{eq:joint_good_data}
\end{equation}
The sample minimizer is then unique. Below fix any data sequence in this probability-one set, also satisfying the query and estimating-equation uniform laws. The conclusions on these sequences imply the asserted convergence in probability and do not condition the privacy guarantee on a statistical event.

\paragraph{Conditional density and moments.}
Let $t_\theta=n\epsilon_{\theta,n}$ and $v=t_\theta(\theta-\hat\theta)$. The unnormalized density of the scaled parameter is
\[
 p_n^{\rm un}(v)=\exp\left\{-\frac{\|t_\theta g_n(\hat\theta+v/t_\theta)\|}{4C_\phi M}\right\}
 1\{v\in t_\theta(\Theta-\hat\theta)\}.
\]
For each fixed $v$ the support indicator equals one eventually, by \eqref{eq:joint_good_data} and $t_\theta\to\infty$. Taylor's theorem with the bounded third derivatives gives $t_\theta g_n(\hat\theta+v/t_\theta)\to Jv$. Strong convexity gives the global bound
\[
 0\leq p_n^{\rm un}(v)\leq\exp\{-\kappa\|v\|/(4C_\phi M)\}.
\]
For sufficiently large $n$, the exponent on the unit ball is at least $-\Lambda\|v\|/(4C_\phi M)$, so integration over this ball bounds $\int p_n^{\rm un}$ away from zero. Dominated convergence, also after multiplication by $1+\|v\|^q$, proves convergence of the normalizers and \eqref{eq:kng_weighted_density_limit}. The transformation $v=4C_\phi M J^{-1}w$ identifies the limit as claimed. In particular, every fixed conditional moment converges, and for fixed constants $K_0,b_{\rm tail,0}>0$,
\begin{equation}
 \Pr\{\|t_\theta(\tilde\theta-\hat\theta)\|>u\mid D\}
 \leq K_0 \exp(-b_{\rm tail,0}u)\quad(u\geq0)
 \label{eq:joint_conditional_tail}
\end{equation}
on the fixed sequence for all sufficiently large $n$. A polynomial factor from the radial integral is absorbed by decreasing $b_{\rm tail,0}$.

\paragraph{The protected contrast in conditional squared mean.}
Set $t=n\epsilon_n$, $\Delta_\theta=\tilde\theta-\hat\theta$, and $\nu_j=M L_j/(t\pi_j)$. The $L_j$ are independent of the KNG draw conditional on $D$. Let $\hat B_z=B_z(\hat\theta)$ and $\hat\mu_z=A_z(\hat\theta)/\hat B_z$, which are well defined eventually. Uniform laws give $\hat B_z\to H$ and $\hat\mu_z\to\mu_z$. The margins in Assumption~\ref{ass:fixed-joint} and bounded query derivatives supply a fixed neighborhood of $(\hat\theta,0)$ on which all perturbed denominators exceed $b$ and both perturbed ratios lie strictly in $(0,1)$. On that neighborhood the protected estimator is the smooth function
\[
 T_n^{\rm loc}(\theta,v)=\frac{A_1(\theta)+v_1}{B_1(\theta)+v_2}
 -\frac{A_0(\theta)+v_3}{B_0(\theta)+v_4},
\]
whose second derivatives are uniformly bounded. Consequently, for a fixed sufficiently small $r_1>0$, on $\mathcal N_n=\{\|\Delta_\theta\|+\|\nu\|\leq r_1\}$,
\begin{align*}
 \tilde\tau_b-\hat\tau_b
 &=\gamma_n^\top\Delta_\theta+\ell_{Q,n}^\top\nu+\mathcal R_n,\\
 \gamma_n&=\nabla_\theta T_b(\hat\theta)\longrightarrow\gamma_\tau,\\
 \ell_{Q,n}&=(\hat B_1^{-1},-\hat\mu_1\hat B_1^{-1},
 -\hat B_0^{-1},\hat\mu_0\hat B_0^{-1})^\top,\\
 |\mathcal R_n|&\leq K(\|\Delta_\theta\|+\|\nu\|)^2.
\end{align*}
The formula for $\gamma_\tau$ follows by differentiating the two ratios and taking the uniform-law limit. This joint Taylor expansion includes the first-stage, ratio, and mixed remainders.

Define $Z_n=t(\tilde\tau_b-\hat\tau_b)$ and $Z_n'=t(\gamma_n^\top\Delta_\theta+\ell_{Q,n}^\top\nu)$ on all mechanism outcomes. Equation~\eqref{eq:joint_conditional_tail}, fixed positive shares, and the Laplace tails give
\[
 \Pr(\mathcal N_n^c\mid D)\leq K_1\exp(-b_{\rm tail,1}t),\qquad
 \mathbb E\{(\|t\Delta_\theta\|+\|t\nu\|)^4\mid D\}\leq K_2
\]
eventually on the fixed data sequence. Thus the contribution of $\mathcal N_n$ to $\mathbb E\{(Z_n-Z_n')^2\mid D\}$ is at most $K^2K_2/t^2$. On its complement, $|\tilde\tau_b-\hat\tau_b|\leq2$, while $\mathbb E(|Z_n'|^4\mid D)$ is bounded. Cauchy--Schwarz therefore bounds the complementary contribution by
\[
 8t^2\Pr(\mathcal N_n^c\mid D)
 +2\{\mathbb E(|Z_n'|^4\mid D)\}^{1/2}
      \Pr(\mathcal N_n^c\mid D)^{1/2}\longrightarrow0.
\]
We have proved
\begin{equation}
 \mathbb E\{(Z_n-Z_n')^2\mid D\}\longrightarrow0.
 \label{eq:joint_conditional_remainder}
\end{equation}
All tail probabilities here are conditional on the same database. No unconditional probability for an atypical database has been substituted into a conditional moment bound. In particular, \eqref{eq:joint_conditional_remainder} proves negligibility of the floor and both projections on the required squared-error scale without taking moments of an unprotected noisy ratio.

Weighted density convergence, independence of the four Laplace variables, and convergence of $\gamma_n,\ell_{Q,n}$ imply the conditional weak limit $Z_n'\to L_\pi$. They also give conditional second-moment convergence. The cross term between its KNG and Laplace parts has exactly zero conditional expectation because the Laplace part is centered and independent. For the radial law, polar integration gives $\mathbb E(W_dW_d^\top)=\{\Gamma(d+2)/(d\Gamma(d))\}I_d=(d+1)I_d$. The limiting second moment is consequently \eqref{eq:joint_cost_coefficients} summed with the squared inverse shares. Equation~\eqref{eq:joint_conditional_remainder} transfers this conclusion and the conditional weak limit from $Z_n'$ to $Z_n$.

\paragraph{Sampling law and independence.}
On the same typical data sequences the nonprivate stacked estimate $\hat\xi=(\hat\theta^\top,\hat\mu_1,\hat\mu_0)^\top$ solves the empirical equations \eqref{eq:stacked-score} and is consistent for $\xi_0$. The population Jacobian has the nonsingular block form
\[
 J_\psi=\begin{pmatrix}
 -J&0&0\\ C_1^\top&-H&0\\ C_0^\top&0&-H
 \end{pmatrix}.
\]
A Taylor expansion of the empirical equations along the segment between $\hat\xi$ and $\xi_0$, with the uniform Jacobian law, yields
\[
 \sqrt n(\hat\xi-\xi_0)
 =-J_\psi^{-1}\frac1{\sqrt n}\sum_{i=1}^n\psi(O_i;\xi_0)+o_p(1).
\]
The summands have mean zero by identification and correct specification and are bounded. The multivariate central limit theorem proves the sampling limit with the sandwich $V$ in \eqref{eq:sandwich}. The contrast influence function is
\[
 \frac{u_1(O;\xi_0)-u_0(O;\xi_0)}{H}
 +\gamma_\tau^\top J^{-1}s(O;\theta_0).
\]
The nonprivate protections are inactive with probability tending to one, so this expansion applies to $\hat\tau_b$.

For any bounded continuous $f$, conditional convergence proved above gives $\mathbb E\{f(Z_n)\mid D\}\to_p\mathbb E f(L_\pi)$. This convergence is also in $L^1$ because the conditional expectations are bounded. Multiplying by any bounded continuous function of $\sqrt n(\hat\tau_b-\tau_h)$ and taking expectations therefore factorizes the joint limit. This proves \eqref{eq:joint_sampling_privacy_limit} with independent components. Finally,
\[
 \sqrt n(\tilde\tau_b-\tau_h)
 =\sqrt n(\hat\tau_b-\tau_h)+\frac{Z_n}{\sqrt n\epsilon_n},
 \qquad
 n\epsilon_n(\tilde\tau_b-\tau_h)
 =\sqrt n\epsilon_n\sqrt n(\hat\tau_b-\tau_h)+Z_n
\]
give the three stated regimes.
\end{proof}

\begin{proof}[Proof of Corollary~\ref{cor:oracle_allocation}]
H\"older's inequality gives
\[
 \sum_{k=0}^4a_k^{1/3}
 =\sum_{k=0}^4(a_k/\pi_k^2)^{1/3}\pi_k^{2/3}
 \leq C(\pi)^{1/3}\left(\sum_{k=0}^4\pi_k\right)^{2/3}.
\]
Equality requires $a_k/\pi_k^2$ proportional to $\pi_k$. If all coefficients are positive this yields the unique displayed solution. Under the interior arm assumptions, $a_1,\ldots,a_4>0$. If $a_0=0$, equality would require $\pi_0=0$, and strictly positive shares can approach but cannot attain the displayed infimum. With a positive public floor the feasible simplex is compact. The optimality conditions for its convex objective give $-2a_k/\pi_k^3+\lambda=0$ on coordinates above the floor and $2a_k/\underline\pi^3\leq\lambda$ at the floor. These conditions give the stated formula. Its normalization has a unique positive solution since its left-hand side decreases continuously from infinity to $5\underline\pi<1$ and is strictly decreasing wherever it exceeds $5\underline\pi$.
\end{proof}

% End section formerly in supplement/joint_limit_proofs.tex
% Begin section formerly in supplement/inference_proofs.tex

\section{Proof of private full-noise inference}
\label{app:inference-proofs}

\paragraph{Joint equations and influence function.}
The nonzero derivatives are
\begin{align}
 \partial_\theta s
 &=\{Zw_1(e)k_1(e)-(1-Z)w_0(e)k_0(e)\}\phi(X)\phi(X)^\top,\nonumber\\
 \partial_\theta u_z&=u_zk_z(e)\phi(X)^\top,\qquad
 \partial_{\mu_z}u_z=-1\{Z=z\}w_z(e).
 \label{eq:sandwich-derivatives}
\end{align}
Conditional expectation and correct propensity specification give
$\E(s\mid X)=0$ and
$\E(\partial_\theta s\mid X)=-h_0(X)\phi(X)\phi(X)^\top$.
Thus, with $m_z(X)=\E\{Y(z)\mid X\}$,
\begin{equation}
 J_\psi=\begin{pmatrix}-J&0&0\\C_1^\top&-H&0\\C_0^\top&0&-H\end{pmatrix},\quad
 J=\E(h_0\phi\phi^\top),\quad
 C_z=\E[\,h_0\{m_z(X)-\mu_z\}k_z(e_{\theta_0})\phi(X)\,].
 \label{eq:population-bread}
\end{equation}
The empirical-equation expansion proved in
Appendix~\ref{app:joint_privacy_proofs} gives
$\operatorname{IF}_\theta=J^{-1}s$ and
$\operatorname{IF}_{\mu_z}=\{u_z+C_z^\top J^{-1}s\}/H$.
Their difference proves \eqref{eq:wate-influence-function} and
\eqref{eq:sandwich}, including the propensity-estimation contribution.
For $\tilde J_\psi=U_{\mathrm{jac}}\operatorname{diag}(\sigma_j)V_{\mathrm{jac}}^\top$, the regularized inverse is
\[
\tilde J_{\psi,\zeta}^{-1}
=V_{\mathrm{jac}}\operatorname{diag}\{\max(\sigma_j,\zeta_n)^{-1}\}U_{\mathrm{jac}}^\top.
\]

\paragraph{Privacy and coefficient consistency.}
Fix any previous output $(\tilde\theta,\tilde Q)$. Then $\tilde\xi$ is
fixed, $\tilde\theta\in\Theta$, and $\tilde\mu_z\in[0,1]$.
Since $|k_z|\le1$ and each record contributes to only one arm,
\[
 \|\psi(O;\tilde\xi)\|_1\le M(C_{\phi,1}+1),\qquad
 \sum_{j,k}|\partial_{\xi_k}\psi_j(O;\tilde\xi)|
 \le M(C_{\phi,1}^2+C_{\phi,1}+1).
\]
The entrywise absolute sum of $\psi\psi^\top$ is
$\|\psi\|_1^2\le M^2(C_{\phi,1}+1)^2$. Hence the vectorized matrix averages have
replacement sensitivities $2M(C_{\phi,1}^2+C_{\phi,1}+1)/n$ and $2M^2(C_{\phi,1}+1)^2/n$. The Laplace releases cost
$\epsilon_{\mathrm{jac}}$ and $\epsilon_{\mathrm{mom}}$ for every fixed prior output, including on
empty-arm or singular databases. Adaptive composition and post-processing
prove privacy without statistical regularity conditions.

The joint privacy theorem, fixed positive shares, and $n\epsilon_n\to\infty$
give $\tilde\theta\to_p\theta_0$ and vanishing query noise. Uniform laws
for the bounded smooth weights, $H>b$, and the armwise interior means then
give $\tilde\xi\to_p\xi_0$ and $\tilde H_z\to_p H$.
The entries of $\partial_\xi\psi$ and $\psi\psi^\top$ are bounded and
uniformly Lipschitz on $\Theta\times[0,1]^2$, so their sample averages at
$\tilde\xi$ converge to $J_\psi$ and $\Sigma_\psi$ even though $\tilde\xi$ depends on the
sample. Fixed dimension and $n\epsilon_{\mathrm{jac}},n\epsilon_{\mathrm{mom}}\to\infty$ make the
added matrix noises vanish. Symmetrization and positive semidefinite
projection are continuous. Nonsingularity of $J_\psi$, positive definiteness of
$J$, and $\zeta_n\to0$ make both inverse floors inactive with probability
tending to one. Continuous mapping proves all coefficient limits in
Theorem~\ref{thm:private-inference}.

\paragraph{Reference quantiles and coverage.}
Set
\[
 t_n=\left\{\frac1n+\frac1{n^2\epsilon_n^2}\right\}^{1/2},\qquad
 u_n=\frac{n^{-1/2}}{t_n},\qquad
 v_n=\frac{(n\epsilon_n)^{-1}}{t_n}.
\]
As $\sqrt n\epsilon_n$ tends to $\infty$, $c_\epsilon\in(0,\infty)$, or $0$,
$(u_n,v_n)$ tends respectively to $(1,0)$, $(c_\epsilon,1)/\sqrt{1+c_\epsilon^2}$, or
$(0,1)$. Denote this limit by $(u,v)$. The joint limit and its independence
conclusion give
\[
 (\tilde\tau_b-\tau_h)/t_n\Longrightarrow
 Z_{u,v}:=uG_V+vL_\pi.
\]
Conditional on the transcript, $\widehat U_n/t_n$ replaces the population
coefficients of this reference law by consistent private estimates.
Along every subsequence where these coefficients converge almost surely,
using common independent reference draws proves weak convergence of the
conditional laws to $Z_{u,v}$. Thus this convergence holds in probability.

The limit is symmetric with a continuous, strictly positive density.
For $u>0$, convolution with the nondegenerate Gaussian establishes this
property. For $u=0$, the independent numerator term $ML_1/(H\pi_1)$ is
nondegenerate Laplace and gives the same conclusion. Therefore
$|Z_{u,v}|$ has a continuous, strictly increasing distribution on
$[0,\infty)$. Conditional weak convergence implies uniform convergence
in probability of the absolute-value distribution functions and convergence
of their quantiles. If $q_{u,v}$ is the limiting $(1-\alpha_{\rm CI})$
quantile, then
\[
 q_{1-\alpha_{\rm CI},n}/t_n\to_p q_{u,v},\qquad
 \Pr\{|\tilde\tau_b-\tau_h|\le q_{1-\alpha_{\rm CI},n}\}
 \longrightarrow1-\alpha_{\rm CI}.
\]
Intersecting with $[-1,1]$ preserves coverage because $\tau_h$ belongs to
that range.

For simulation, draw $W_d=R_WU_W$ with independent
$R_W\sim\operatorname{Gamma}(d,1)$ and $U_W$ uniform on the unit sphere.
For $N_{{\rm ref},n}$ independent reference draws, let $\widehat F_{N_{{\rm ref},n}}$ and $F_n$
be the empirical and conditional distribution functions of
$|\widehat U_n|$. The sharp Dvoretzky--Kiefer--Wolfowitz inequality
\citep{Massart1990} gives, for every $u>0$,
\begin{equation}
 \Pr\!\left\{\sup_t|\widehat F_{N_{{\rm ref},n}}(t)-F_n(t)|>u
                  \,\middle|\,\text{releases}\right\}
 \le2\exp(-2N_{{\rm ref},n}u^2).
 \label{eq:reference-monte-carlo}
\end{equation}
Thus $N_{{\rm ref},n}\to\infty$ and the same strict-quantile argument prove coverage
for empirical quantiles. On the complementary event in
\eqref{eq:reference-monte-carlo}, the empirical quantile at
$1-\alpha_{\rm CI}+u$ bounds the exact reference quantile from above.
Return $[-1,1]$ if this level reaches one. Choosing $u\to0$ with
$N_{{\rm ref},n}u^2\to\infty$ preserves asymptotic coverage. This calibration controls
simulation error, not finite-sample causal coverage.

\paragraph{A vanishing budget fraction for private moments.}
Only consistency is required of the private moments. For bounded total
budgets with $n\epsilon_{\mathrm{total},n}\to\infty$, set
\[
 r_n=\min\{1/2,(n\epsilon_{\mathrm{total},n})^{-1/2}\},\qquad
 \epsilon_n=(1-r_n)\epsilon_{\mathrm{total},n}.
\]
Assign fixed positive shares of $r_n\epsilon_{\mathrm{total},n}$ to
$\epsilon_{\mathrm{jac}}$ and $\epsilon_{\mathrm{mom}}$. Then $r_n\to0$ and
$n\epsilon_{\mathrm{jac}},n\epsilon_{\mathrm{mom}}\to\infty$. Relative to spending the full total
budget on the point mechanism, its leading conditional cost is multiplied
by $(1-r_n)^{-2}\to1$. This feasible allocation does not claim to optimize
finite-sample interval length.

\section{Rejection sampling on the public domain}
\label{app:sampling}
With $a=n\epsilon_\theta/(4C_\phi M)$, uniform rejection accepts with
probability $\exp\{-a\|g_n(\theta)\|_2\}$ on $\Theta$.
Compactness and continuity give positive acceptance probability on every
database, including empty arms and singular designs. The experiments use
the public Euclidean ball of radius
$R_\Theta=\log\{(1-\eta)/\eta\}/C_\phi$, centered at zero. Cauchy--Schwarz
establishes the propensity restriction on the full record domain.

\paragraph{Certified shell envelope.}
For a numerical center $\theta_c\in\Theta$, the residual is $r_c=\|g_n(\theta_c)\|_2$.
If $\nabla^2L_n(\theta)\succeq\kappa I_d$ throughout $\Theta$, strong
monotonicity gives
$\|g_n(\theta)\|_2\ge\kappa\|\theta-\theta_c\|_2-r_c$.
The implemented refinement also uses the row Hessian
$c_z\{\operatorname{expit}(v)\}\phi\phi^\top$, where $v=\theta^\top\phi$. Up to a common weight
scale, the nonzero functions are $\exp(-v),\exp(v)$ for ATE,
$\operatorname{expit}(v)\{1-\operatorname{expit}(v)\}$ for ATO, $\exp(v)$ in the control arm for ATT, and $\exp(-v)$
in the treated arm for ATC. Each satisfies
$|\partial_v\log c_z\{\operatorname{expit}(v)\}|\le1$, and zero contributions remain zero.
For a unit vector $u$ and $\lambda_c=\lambda_{\min}\{\nabla^2L_n(\theta_c)\}$,
integration along the segment gives
\[
 \nabla^2L_n(\theta_c+tu)\succeq \exp(-C_\phi t)\nabla^2L_n(\theta_c),\qquad
 u^\top\{g_n(\theta_c+ru)-g_n(\theta_c)\}
 \ge\frac{\lambda_c}{C_\phi}(1-\exp(-C_\phi r)).
\]
Consequently the monotone lower envelope
\begin{equation}\label{eq:shell-bound}
 \ell_{\rm env}(r)=\max\left[0,\kappa r-r_c,
 \frac{\lambda_c}{C_\phi}(1-\exp(-C_\phi r))-r_c\right]
,\qquad
 \|g_n(\theta)\|_2\ge \ell_{\rm env}(\|\theta-\theta_c\|_2).
\end{equation}
This directional argument does not interchange noncommuting matrices.
Neither a local Hessian nor an approximate root permits dropping the
exponential factor or the center residual.

For shell boundaries $0=r_0^{\rm sh}<\cdots<r_K^{\rm sh}=R_\Theta+\|\theta_c\|_2$, let
$\varrho_k=\exp\{-a \ell_{\rm env}(r_{k-1}^{\rm sh})\}$. The sampler selects shell $k$ with probability
proportional to $\varrho_k((r_k^{\rm sh})^d-(r_{k-1}^{\rm sh})^d)$, then draws a uniform direction and
$r^d$ uniformly on $[(r_{k-1}^{\rm sh})^d,(r_k^{\rm sh})^d]$. The proposal density is
proportional to $\varrho_k$ on that shell. Proposals outside $\Theta$ are rejected
and those inside are accepted with probability
\[
 \exp[-a\{\|g_n(\theta)\|_2-\ell_{\rm env}(r_{k-1}^{\rm sh})\}].
\]
Monotonicity of $\ell_{\rm env}$ makes this at most one. Multiplying by the proposal
density cancels $\varrho_k$ and gives the KNG target exactly. The center, shell
locations, and certified constants may therefore depend internally on the
database without changing the released law. We use geometrically spaced
shells. If $\kappa=\lambda_c=0$, this reduces to uniform rejection on an
enclosing ball.

For $G_D=n^{-1}\sum_i\phi_i\phi_i^\top$, the global certificates are
$\kappa=\{\eta/(1-\eta)\}\lambda_{\min}(G_D)$ for ATE and
$\kappa=\eta(1-\eta)\lambda_{\min}(G_D)$ for ATO.
ATT and ATC use $\eta/(1-\eta)$ times the minimum eigenvalue of the control-arm
and treated-arm Gram matrices, respectively, on the same average scale.
Common weight scaling multiplies these certificates accordingly. A center
just outside the ball is projected into it. No score-root or score-norm
optimality property of the center is required.

\paragraph{Uniform-envelope comparison and implementation.}
The experiments choose the smaller of the shell envelope integral and a
certified uniform envelope on $\Theta$. For each record,
$v_i\in[-R_\Theta\|\phi_i\|_2,R_\Theta\|\phi_i\|_2]$, and its scalar negative-score
contribution is nondecreasing in $v_i$. Evaluating both endpoints,
multiplying by each feature coordinate with the appropriate sign, and
averaging produces intervals $[\ell_j,u_j]$ containing all coordinates of
$g_n(\Theta)$. Hence
\[
 m_D^2=\sum_j\{\max(\ell_j,0)+\min(u_j,0)\}^2
 \le\inf_{\theta\in\Theta}\|g_n(\theta)\|_2^2.
\]
Uniform proposals accept with probability
$\exp[-a\{\|g_n(\theta)\|_2-m_D\}]$, whose common envelope factor cancels.
Known ball and shell volumes give the envelope-integral comparison
analytically. All proposal choices remain internal.

Numerical checks cover all four estimands, approximate and constrained
centers, absent arms, singular designs, small budgets, and agreement with
independent two-dimensional integration. The nonprivate comparator is
checked separately by its projected-gradient residual.

Exactness assumes real arithmetic and exact random draws. The mathematical
mechanism is uncapped and returns only its declared private transcript.
Public and synthetic experiments impose proposal caps to measure feasibility
and retain every failure. Such caps or failure messages are not private
implementations of the uncapped mechanism. Finite precision and the
random-number generator are outside the proof. Observable proposal counts
and running times should be hidden or protected separately \citep{Awan2024}.

% End section formerly in supplement/sampling.tex
% Begin section formerly in supplement/empirical_details.tex

\section{Simulation and Data Analysis Details}
\label{sec:simulation_details}
\label{app:empirical-checks}
\subsection{Generating Laws and Computational Protocol}
\label{app:empirical-designs}
In the multivariate simulation, $X_1,\ldots,X_4$ are independent and uniform on $\{-1,0,1\}$, with
\begin{align*}
 \operatorname{logit}e_0(X)&=-0.1+0.27X_1-0.23X_2+0.20X_3+0.16X_4,\\
 \operatorname{logit}m_0(X)&=-0.8+0.7X_1-0.6X_2+0.45X_3+0.35X_4
                         +0.35X_1X_2-0.25X_3^2,\\
 \operatorname{logit}m_1(X)&=\operatorname{logit}m_0(X)+0.65+0.25X_2-0.2X_4+0.2X_1X_3.
\end{align*}
Potential outcomes are Bernoulli and conditionally independent of treatment. Propensities range over $[0.2769,0.6814]$, with coefficient norm $0.4488<0.6200$, the public radius. The CBSR variance coefficient is $V=0.85586$ and the contrast-derivative norm is 0.1302.

Table~\ref{tab:correct-model-comparison} uses 1,000 replicated datasets from this law, shared across all six methods. The two nonprivate fits use their own stacked sandwich variances. The private CBSR and logistic fits use the same public feature map, domain, queries, and budget shares, with their respective gradient sensitivities. Both private intervals use the full reference distribution in \eqref{eq:full-noise-reference}. The Lee and Guha--Reiter tuning and interval construction are unchanged from Appendix~\ref{app:empirical-comparators}.

For allocation and shrinking-budget studies, $X$ is uniform on eight equally spaced points in $[-1,1]$, $\phi=(1,X)^\top$, and
\begin{align*}
 e_0(X)&=\operatorname{expit}(-0.1+0.6X),\\
 m_0(X)&=0.32+0.27X+0.10\{X^2-E(X^2)\},\\
 m_1(X)&=m_0(X)+0.12+0.10X.
\end{align*}
The binary potential outcomes satisfy conditional treatment independence. The ATE is 0.12, $C_\phi=\sqrt2$, and the true propensity coefficient is interior. Population evaluation gives
\begin{align*}
 \gamma_\tau&=(0.00627,-0.13814)^\top,\qquad V=0.79149,\\
 (a_0,a_1,a_2,a_3,a_4)&=(249.46,50,9.68,50,5.12).
\end{align*}
The derivative agrees with centered finite differences. Allocation releases spend 0.8 on point estimation only, and empirical costs include the cross term between stages.

\subsection{Additional Correctly Specified Results}
\label{app:correct-budget-diagnostics}

For the configurations in Table~\ref{tab:simulation_well_specified}, coverage at $n=10^5$ is close to 0.95 and the intervals are informative. At budget 1, omitting the KNG contribution reduces coverage from 0.956 to 0.914, and retaining only the sandwich sampling term reduces it to 0.902. At budget 5, the private mean width 0.0115 is close to the nonprivate width 0.0115. The smaller sample at budget 1 is uninformative despite coverage 1.000. In that configuration, 24.6\% of intervals equal $[-1,1]$, and private bread and Hessian regularization activate in 40.7\% and 88.1\% of runs. Positive-definite regularization ensures a defined calculation but does not prove conservative coverage.

\begin{table}[htbp]
\centering\small
\caption{Method comparison under the correctly specified multivariate propensity model with $n=10^5$, using 1,000 datasets paired across methods. NP is nonprivate. DP budgets are 1.6 for points and 0.4 for intervals. Lee is $(1.6,10^{-6})$-DP without intervals. Other complete releases are pure 2-DP. Intervals have nominal 95\% coverage and are intersected with $[-1,1]$. Balance is the norm of arm-normalized feature-mean differences. $\dagger$ denotes the Guha--Reiter interval adaptation in Appendix~\ref{app:empirical-comparators}. All attempts completed.}
\label{tab:correct-model-comparison}
\setlength{\tabcolsep}{5pt}
\begin{tabular}{lrrrrr}
\toprule
Estimator & Bias & RMSE & Balance & Coverage & Mean length\\
\midrule
CBSR, NP & 0.0000 & 0.0029 & 0.0000 & 0.946 & 0.0115\\
CBSR, DP & 0.0000 & 0.0030 & 0.0060 & 0.937 & 0.0119\\
Logistic, NP & 0.0000 & 0.0029 & 0.0004 & 0.948 & 0.0115\\
KNG-logistic, DP & 0.0000 & 0.0029 & 0.0043 & 0.945 & 0.0124\\
Lee et al., DP & 0.0161 & 0.0171 & -- & -- & --\\
Guha--Reiter$^\dagger$, DP & -0.0004 & 0.0181 & -- & 0.950 & 0.0756\\
\bottomrule
\end{tabular}
\end{table}

Table~\ref{tab:correct-model-comparison} shows negligible bias for CBSR and logistic weighting, with RMSE between 0.0029 and 0.0030. Nonprivate CBSR achieves balance to the displayed precision. Its private version has imbalance 0.0060, compared with 0.0043 for KNG-logistic. Private CBSR and KNG-logistic have coverage 0.937 and 0.945, with mean interval lengths 0.0119 and 0.0124, respectively. Lee and Guha--Reiter have larger RMSE, 0.0171 and 0.0181. Guha--Reiter achieves coverage 0.950 with substantially wider intervals, of mean length 0.0756.

\subsection{Shrinking Budgets and Inference Diagnostics}
\label{app:interval-diagnostics}
For each $n\in\{10^4,10^6,10^8\}$ we generate 1,000 independent two-dimensional databases and pair them across the total-budget sequences
\[
 1,\qquad 30n^{-1/2},\qquad 300n^{-3/4}.
\]
Point budgets are 0.8 times these values, so $\sqrt n\epsilon_{{\rm point},n}$ tends to infinity, equals 24, or tends to zero, respectively. Even in the last regime, $n\epsilon_{{\rm point},n}$ diverges. Moment budgets satisfy consistency. The 9,000 analyses use 3,000 independent datasets. Critical and low sequences coincide at $n=10^4$.

\begin{table}[t]
\centering\small
\caption{Superpopulation 95\% coverage and mean interval width, using 1,000 databases per row. Ablations share the private point estimate and moment transcript. All 9,000 analyses completed without point flooring or arm projection.}
\label{tab:r3-coverage}
\begin{tabular}{llrrrrr}
\toprule
$n$ & Sequence & $\epsilon_{\rm total}$ & Full & Omit KNG & Wald & Width\\
\midrule
$10^4$ & Fixed & 1 & 0.959 & 0.900 & 0.815 & 0.0808\\
$10^4$ & Critical & 0.3 & 0.992 & 0.922 & 0.771 & 0.6786\\
$10^6$ & Fixed & 1 & 0.945 & 0.945 & 0.944 & 0.003517\\
$10^6$ & Critical & 0.03 & 0.938 & 0.793 & 0.374 & 0.01672\\
$10^6$ & Low & 0.009487 & 0.930 & 0.775 & 0.172 & 0.06937\\
$10^8$ & Fixed & 1 & 0.935 & 0.936 & 0.936 & 0.0003486\\
$10^8$ & Critical & 0.003 & 0.957 & 0.787 & 0.354 & 0.001642\\
$10^8$ & Low & 0.0003 & 0.939 & 0.770 & 0.044 & 0.01643\\
\bottomrule
\end{tabular}
\end{table}

At $n=10^6$ under the low budget, replacing only the sampling variance by its population value changes coverage from 0.930 to 0.929. Replacing only the privacy-law coefficients gives 0.949. This implicates private coefficient estimation but does not repair operational coverage, so this configuration is excluded from the empirically supported range for nominal inference. Mean released variance is $2.172>0.79149$, showing that upward average variance error need not ensure conservative coverage. 

\subsection{Magnitude of the Explicit Conditional Bound}
The finite-sample bound is evaluated on the same 60 databases as the allocation study. For the ATE score and public ball, valid global constants are
\[
 \kappa_D=\frac{\eta}{1-\eta}\lambda_{\min}(G_D),\qquad
 \Lambda_D=\frac{1-\eta}{\eta}\lambda_{\max}(G_D),\qquad
 r_{0,D}=R_\Theta-\|\hat\theta\|_2.
\]
Here $R_\Theta=\log\{(1-\eta)/\eta\}/C_\phi$ is the public ball radius from Appendix~\ref{app:sampling}.
Arm denominators are uniformly at least $\rho_D=\min(n_0,n_1)/\{n(1-\eta)\}$. We use the inactive-floor Lipschitz constant when $\rho_D\geq b$, the exact gamma distribution function, and average-query noise scales. All roots are interior with score residual below $10^{-9}$.

\begin{table}[t]
\centering\small
\caption{Unscaled conditional squared cost and explicit bound. Costs average draws within each of 20 databases, then across databases. Parentheses give Monte Carlo standard errors across databases. Nontrivial bounds are below 4.}
\label{tab:r3-finite-bound}
\label{tab:r3-cost}
\begin{tabular}{rlrrr}
\toprule
$n$ & Allocation & Cost (MCSE) & Mean bound & Nontrivial\\
\midrule
$10^4$ & Equal & $1.38\times10^{-4}$ ($3.74\times10^{-6}$) & $4.00$ & 0\%\\
$10^4$ & Oracle & $8.37\times10^{-5}$ ($2.69\times10^{-6}$) & $4.00$ & 0\%\\
$10^6$ & Equal & $1.36\times10^{-8}$ ($4.97\times10^{-10}$) & $2.17\times10^{-3}$ & 100\%\\
$10^6$ & Oracle & $8.39\times10^{-9}$ ($2.46\times10^{-10}$) & $7.35\times10^{-4}$ & 100\%\\
$10^8$ & Equal & $1.36\times10^{-12}$ ($3.75\times10^{-14}$) & $2.17\times10^{-7}$ & 100\%\\
$10^8$ & Oracle & $8.60\times10^{-13}$ ($2.74\times10^{-14}$) & $7.35\times10^{-8}$ & 100\%\\
\bottomrule
\end{tabular}
\end{table}

The bound is trivial at the smallest sample and conservative elsewhere. Its constants cover the public domain, while the local expansion better describes allocation at these interior fits.

\subsection{Model Misspecification and Comparators}
\label{app:empirical-comparators}
On the common eight-point support, potential outcomes are binary and conditionally independent of treatment, with $m_1(X)=m_0(X)+0.12$. The generating laws are
\[
 \begin{aligned}
 \text{Linear}\quad &\operatorname{logit}e_0(X)=-0.3+0.8X+1.7(X^2-3/7),\\
 &m_0(X)=0.32+0.27X,\\
 \text{Quadratic}\quad &\operatorname{logit}e_0(X)=-0.4+0.2X+0.3(X^2-3/7)+3(X^3-37X/49),\\
 &m_0(X)=0.32+0.20X+0.20(X^2-3/7).
 \end{aligned}
\]

KNG uses $(\eta,b)=(0.2,0.1)$ and five equal point-budget shares. All Table~\ref{tab:simulation_misspecified} releases completed. The quadratic paired CBSR-minus-logistic MSE difference is $-1.63\times10^{-6}$ (Monte Carlo standard error $2.26\times10^{-7}$). Nonprivate intervals use the joint sandwich. Private intervals retain the point estimates, adding 0.2 per moment matrix and 20,000 reference draws. Quadratic CBSR coverage has Monte Carlo standard error 0.0052. Quadratic logistic bread and Hessian flooring each activate in 99.8\% of runs, versus neither for CBSR.

The Lee and Guha--Reiter settings below apply to both Tables~\ref{tab:correct-model-comparison} and \ref{tab:simulation_misspecified}. 
The \citet{Lee2019} fit uses $\phi(X)/\sqrt d$, a random half-sample, and ridge penalty $\lambda_{\rm Lee}\|\theta\|^2/2$ including the intercept, with their default $\lambda_{\rm Lee}=0.1$. Coefficient sensitivity is $2/(n_{\rm train}\lambda_{\rm Lee})$. Their Appendix B Horvitz--Thompson contrast uses the remaining sample, denominator floors $\max(\eta_{\rm cmp},\tilde e)$ and $\max(\eta_{\rm cmp},1-\tilde e)$ with $\eta_{\rm cmp}=0.05$, and sensitivity $2/(n_{\rm est}\eta_{\rm cmp})$. Analytic Gaussian calibration \citep{Balle2018} replaces their $\epsilon<1$ formula, giving standard deviations 2.736655 times sensitivity for $(1.6,10^{-6})$-DP by parallel composition.

Guha--Reiter uses $K_{\rm GR}=100$ equal partitions, logistic ridge $10^{-8}$, propensities truncated to $[0.05,0.95]$, and subgroup H\'ajek estimators with empty-arm means zero. Their average receives $\operatorname{Lap}\{0,2/(K_{\rm GR}\epsilon_{\rm point})\}$ noise and posterior-mean postprocessing under a uniform $[-1,1]$ prior. For intervals, subgroup variances $\hat v_{m,\mathrm{GR}}$ use their Table 1 ATE formula with empirical binomial proportions at each covariate level and arm, or $1/2$ in empty cells. The variance of the average is $\hat v_{\mathrm{GR}}=\sum_m\hat v_{m,\mathrm{GR}}/K_{\rm GR}^2$, bounded by $v_{\max,\mathrm{GR}}=1/(2\eta_{\rm cmp}n)$ with conservative sensitivity $\Delta_{v,\mathrm{GR}}=1/(\eta_{\rm cmp}nK_{\rm GR})$, where $\eta_{\rm cmp}=0.05$. These are unscaled estimator variances, whereas $V$ is the root-$n$ variance coefficient estimated by $\tilde V$, so its contribution to estimator variance is $V/n$. The variance release adds $\operatorname{Lap}(0,\Delta_{v,\mathrm{GR}}/0.4)$ noise. The variance prior is uniform on $[0,v_{\max,\mathrm{GR}}]$. Both bounded Laplace posteriors feed 20,000 normal-mixture draws for equal-tailed intervals. This adaptation resolves inconsistent variance and noise scaling.

\begin{table}[htbp]
\centering\small
\caption{Linear/quadratic results on the Table~\ref{tab:simulation_misspecified} datasets. Lee entries are RMSE. Columns give Lee training fractions or Guha--Reiter partition counts, fixed before evaluation. Guha--Reiter intervals use the stated adaptation.}
\label{tab:comparator-tuning}
\setlength{\tabcolsep}{5pt}
\begin{tabular}{lcccc}
\toprule
Lee ridge $\lambda_{\rm Lee}$ & 0.25 & 0.50 & 0.75 & \\
\midrule
0.001 & 0.0595/0.0584 & 0.0308/0.0281 & 0.0221/0.0210 & \\
0.01 & 0.0104/0.0124 & 0.0097/0.0114 & 0.0111/0.0134 & \\
0.1 & 0.0055/0.0544 & 0.0063/0.0544 & 0.0090/0.0552 & \\
\midrule
Guha--Reiter $K_{\rm GR}$ & 50 & 100 & 200 & 500\\
RMSE & 0.0355/0.0363 & 0.0189/0.0184 & 0.0116/0.0096 & 0.0086/0.0048\\
Coverage & 0.958/0.944 & 0.955/0.942 & 0.917/0.945 & 0.654/0.940\\
Mean length & 0.1504/0.1505 & 0.0757/0.0758 & 0.0389/0.0389 & 0.0182/0.0182\\
\bottomrule
\end{tabular}
\end{table}

An omitted-prognostic check adds $0.10(X^2-3/7)$ to both linear-scenario outcome means, retaining its propensity and basis. In 1,000 runs, private CBSR/logistic RMSEs are 0.0230/0.0162 and imbalances 0.0012/0.0260. Logistic benefits from bias cancellation. Guha--Reiter RMSE decreases from 0.0248 at $K_{\rm GR}=100$ to 0.0163 at $K_{\rm GR}=500$. Thus supplied-feature balance alone does not ensure smaller causal error.

Stress checks use this law with basis $(1,X,X^2,X^3)^\top$ or propensity $\operatorname{expit}(-1.2+2.4X+1.7X^2)$. The former has $C_\phi=2$ and radius $\log(4)/2$, changing domain geometry as well as dimension. The latter retains $(1,X)^\top$. Both fits have active constraints. 

\subsection{Email Campaign Analysis and Calibration}
\label{app:customer-details}
To ensure the boundedness of covariates, we apply rowwise transformations: $2(r-1)/11-1$ for recency $r\in[1,12]$, $2\min(s,1000)/1000-1$ for nonnegative spending, and $2w-1$ for the purchase indicator. The spending cap affects 844 customers, who remain in the analysis. We use $C_\phi=2$ and the same public overlap and floor as in the simulations.

For the email campaign data, the nonprivate CBSR fit has score norm $1.4\times10^{-14}$ and lies strictly inside the public ball. Bread and Hessian regularization activate in 21\% and 28\% of private CBSR draws at budget 1, respectively, compared with neither at budget 5. Repeated privatizations assess conditional privacy error, not sampling coverage, and their collection is not one release at the displayed budget.

Across all 100 CBSR draws at each total budget 1, 5, and 10, mean interval widths are 4.754, 1.292 and 1.260 percentage points, respectively, and their 90th percentiles are 7.472, 1.552 and 1.437. Table~\ref{tab:customer-private} instead displays the interval and width corresponding to its reported point estimate. The balance diagnostics in Table~\ref{tab:customer-balance} use arm-normalized weights before query noise. They are not standardized mean differences and do not constitute additional private releases.

\begin{table}[t]
\centering\small
\caption{Logistic IPW for the email campaign with CBSR's transforms, protected queries, and budget shares. Each private row completed 100 draws. Units and first-draw reporting follow Table~\ref{tab:customer-private}.}
\label{tab:customer-logistic}
\begin{tabular}{lrrrr}
\toprule
$\epsilon_{\rm total}$ & Estimate & Paired RMSE & Mean width & Width 90th pct.\\
\midrule
Nonprivate & 4.519 & -- & 1.261 & --\\
1 & 4.503 & 0.161 & 2.790 & 4.101\\
5 & 4.528 & 0.033 & 1.299 & 1.570\\
10 & 4.509 & 0.017 & 1.265 & 1.411\\
\bottomrule
\end{tabular}
\end{table}
Additional coverage simulations draw iid covariates from the empirical covariate distribution of the email campaign data, with newly generated treatment assignments and binary outcomes. In the order of Section~\ref{sec:analysis},
\begin{align*}
 \operatorname{logit}e_0(x)&=-0.15+0.35x_1+0.25x_2+0.20x_3,\\
 \operatorname{logit}m_0(x)&=-0.6+0.65x_1+0.5x_2+0.35x_3
                         +0.2x_1x_2-0.25x_3^2,\\
 \operatorname{logit}m_1(x)-\operatorname{logit}m_0(x)
                       &=0.5+0.2x_1-0.15x_3+0.1x_1x_3.
\end{align*}

\begingroup
\color{black}\bibliographystyle{Chicago}
\bibliography{literature}

@INPROCEEDINGS{McSherry2007,
  author={McSherry, Frank and Talwar, Kunal},
  booktitle={48th Annual IEEE Symposium on Foundations of Computer Science (FOCS'07)}, 
  title={Mechanism Design via Differential Privacy}, 
  year={2007},
  volume={},
  number={},
  pages={94-103},
  doi={10.1109/FOCS.2007.66}}

@article{Chaudhuri2011,
author = {Chaudhuri, Kamalika and Monteleoni, Claire and Sarwate, Anand D.},
title = {Differentially Private Empirical Risk Minimization},
year = {2011},
issue_date = {2/1/2011},
publisher = {JMLR.org},
volume = {12},
number = {null},
issn = {1532-4435},
journal = {Journal of Machine Learning Research},
month = {jul},
pages = {1069–1109},
numpages = {41}
}

@inproceedings{Nissim2007,
author = {Nissim, Kobbi and Raskhodnikova, Sofya and Smith, Adam},
title = {Smooth sensitivity and sampling in private data analysis},
year = {2007},
isbn = {9781595936318},
publisher = {Association for Computing Machinery},
address = {New York, NY, USA},
url = {https://doi.org/10.1145/1250790.1250803},
doi = {10.1145/1250790.1250803},
booktitle = {Proceedings of the Thirty-Ninth Annual ACM Symposium on Theory of Computing},
pages = {75–84},
numpages = {10},
location = {San Diego, California, USA},
series = {STOC '07}
}

@article{Abadie_Imbens_2006,
author = {Abadie, Alberto and Imbens, Guido W.},
title = {Large Sample Properties of Matching Estimators for Average Treatment Effects},
journal = {Econometrica},
volume = {74},
number = {1},
pages = {235-267},
doi = {https://doi.org/10.1111/j.1468-0262.2006.00655.x},
url = {https://onlinelibrary.wiley.com/doi/abs/10.1111/j.1468-0262.2006.00655.x},
eprint = {https://onlinelibrary.wiley.com/doi/pdf/10.1111/j.1468-0262.2006.00655.x},
year = {2006}
}

@article{Agarwal2021,
  title={Causal Inference with Corrupted Data: Measurement Error, Missing Values, Discretization, and Differential Privacy},
  author={Agarwal, Anish and Singh, Rahul},
  journal={arXiv preprint arXiv:2107.02780},
  year={2021}
}

@article{Awan2024,
 author={Jordan Awan and Vinayak Rao},
 title={Privacy-Aware Rejection Sampling},
 journal={Journal of Machine Learning Research},
 year={2023}, volume={24}, number={74}, pages={1--32},
 url={https://jmlr.org/papers/v24/21-0870.html}
}

@article{Cai2021,
 author={Cai, T. Tony and Wang, Yichen and Zhang, Linjun},
 title={The cost of privacy: Optimal rates of convergence for parameter estimation with differential privacy},
 journal={The Annals of Statistics},
 year={2021}, volume={49}, number={5}, pages={2825--2850},
 doi={10.1214/21-AOS2058}, url={https://arxiv.org/abs/1902.04495}
}

@InProceedings{Chen2024,
  title = 	 {Federated Experiment Design under Distributed Differential Privacy},
  author =       {Chen, Wei-Ning and Cormode, Graham and Bharadwaj, Akash and Romov, Peter and Ozgur, Ayfer},
  booktitle = 	 {Proceedings of The 27th International Conference on Artificial Intelligence and Statistics},
  pages = 	 {2458--2466},
  year = 	 {2024},
  editor = 	 {Dasgupta, Sanjoy and Mandt, Stephan and Li, Yingzhen},
  volume = 	 {238},
  series = 	 {Proceedings of Machine Learning Research},
  month = 	 {02--04 May},
  publisher =    {PMLR},
  url = 	 {https://proceedings.mlr.press/v238/chen24c.html}
}

@article{Chuen2015,
    author = {Chan, Kwun Chuen Gary and Yam, Sheung Chi Phillip and Zhang, Zheng},
    title = "{Globally Efficient Non-Parametric Inference of Average Treatment Effects by Empirical Balancing Calibration Weighting}",
    journal = {Journal of the Royal Statistical Society Series B: Statistical Methodology},
    volume = {78},
    number = {3},
    pages = {673-700},
    year = {2016},
    month = {6},
    issn = {1369-7412},
    doi = {10.1111/rssb.12129},
    url = {https://doi.org/10.1111/rssb.12129},
    eprint = {https://academic.oup.com/jrsssb/article-pdf/78/3/673/49236400/jrsssb\_78\_3\_673.pdf},
}

@article{Dwork2014,
  author = {Dwork, Cynthia and Roth, Aaron},
  ee = {https://www.wikidata.org/entity/Q56623118},
  journal = {Foundations and Trends in Theoretical Computer Science},
  number = {3-4},
  pages = {211-407},
  title = {The Algorithmic Foundations of Differential Privacy.},
  COMMENTurl = {http://dblp.uni-trier.de/db/journals/fttcs/fttcs9.html#DworkR14},
  volume = 9,
  year = 2014
}

@article{Gneiting2007,
author = {Tilmann Gneiting and Adrian E Raftery},
title = {Strictly Proper Scoring Rules, Prediction, and Estimation},
journal = {Journal of the American Statistical Association},
volume = {102},
number = {477},
pages = {359--378},
year = {2007},
publisher = {Taylor \& Francis},
doi = {10.1198/016214506000001437},
URL = {    
        https://doi.org/10.1198/016214506000001437
},
eprint = {     
        https://doi.org/10.1198/016214506000001437
}
}

@article{Hainmueller2012,
   author = {Jens Hainmueller},
   doi = {10.1093/pan/mpr025},
   issn = {10471987},
   issue = {1},
   journal = {Political Analysis},
   pages = {25-46},
   title = {Entropy balancing for causal effects: A multivariate reweighting method to produce balanced samples in observational studies},
   volume = {20},
   year = {2012},
}

@article{Hazlett2020,
   author = {Chad Hazlett},
   doi = {10.5705/ss.202017.0555},
   issn = {10170405},
   issue = {3},
   journal = {Statistica Sinica},
   month = {7},
   pages = {1155-1189},
   publisher = {Institute of Statistical Science},
   title = {Kernel balancing: A flexible non-parametric weighting procedure for estimating causal effects},
   volume = {30},
   year = {2020},
}

@article{Heckman1997,
 ISSN = {00346527, 1467937X},
 URL = {http://www.jstor.org/stable/2971733},
 author = {James J. Heckman and Hidehiko Ichimura and Petra E. Todd},
 journal = {The Review of Economic Studies},
 number = {4},
 pages = {605--654},
 publisher = {[Oxford University Press, Review of Economic Studies, Ltd.]},
 title = {Matching as an Econometric Evaluation Estimator: Evidence from Evaluating a Job Training Programme},
 urldate = {2024-01-12},
 volume = {64},
 year = {1997}
}

@misc{Hillstrom2008Public,
  author = {Hillstrom, Kevin},
  title = {The {MineThatData} E-Mail Analytics and Data Mining Challenge},
  year = {2008},
  howpublished = {MineThatData},
  url = {https://blog.minethatdata.com/2008/03/minethatdata-e-mail-analytics-and-data.html}
}

@article{Huling2024,
url = {https://doi.org/10.1515/jci-2022-0029},
title = {Energy balancing of covariate distributions},
author = {Jared D. Huling and Simon Mak},
pages = {20220029},
volume = {12},
number = {1},
journal = {Journal of Causal Inference},
doi = {doi:10.1515/jci-2022-0029},
year = {2024},
lastchecked = {2024-08-10}
}

@article{Imai2014,
   author = {Kosuke Imai and Marc Ratkovic},
   doi = {10.1111/rssb.12027},
   issn = {13697412},
   issue = {1},
   journal = {Journal of the Royal Statistical Society. Series B: Statistical Methodology},
   month = {1},
   pages = {243-263},
   title = {Covariate balancing propensity score},
   volume = {76},
   year = {2014},
}

@misc{Javanmard2024,
      title={Causal Inference with Differentially Private (Clustered) Outcomes}, 
      author={Adel Javanmard and Vahab Mirrokni and Jean Pouget-Abadie},
      year={2024},
      eprint={2308.00957},
      archivePrefix={arXiv},
      primaryClass={stat.ML},
      url={https://arxiv.org/abs/2308.00957}, 
}

@article{Jianqing2023,
author = {Jianqing Fan and Kosuke Imai and Inbeom Lee and Han Liu and Yang Ning and Xiaolin Yang},
title = {Optimal Covariate Balancing Conditions in Propensity Score Estimation},
journal = {Journal of Business \& Economic Statistics},
volume = {41},
number = {1},
pages = {97--110},
year = {2023},
publisher = {ASA Website},
doi = {10.1080/07350015.2021.2002159},
URL = { 
    
        https://doi.org/10.1080/07350015.2021.2002159
},
eprint = { 
        https://doi.org/10.1080/07350015.2021.2002159
}
}

@article{Kang_Schafer_2007,
author = {Joseph D. Y. Kang and Joseph L. Schafer},
title = {{Demystifying Double Robustness: A Comparison of Alternative Strategies for Estimating a Population Mean from Incomplete Data}},
volume = {22},
journal = {Statistical Science},
number = {4},
publisher = {Institute of Mathematical Statistics},
pages = {523 -- 539},
year = {2007},
doi = {10.1214/07-STS227},
URL = {https://doi.org/10.1214/07-STS227}
}

@article{Komarova2020,
  title={Identification and Formal Privacy Guarantees},
  author={Komarova, Tatiana and Nekipelov, Denis},
  journal={arXiv preprint arXiv:2006.14732},
  year={2020}
}

@inproceedings{Kong0223,
author = {Kong, Insung and Park, Yuha and Jung, Joonhyuk and Lee, Kwonsang and Kim, Yongdai},
title = {Covariate balancing using the integral probability metric for causal inference},
year = {2023},
publisher = {JMLR.org},
booktitle = {Proceedings of the 40th International Conference on Machine Learning},
articleno = {719},
numpages = {32},
location = {Honolulu, Hawaii, USA},
series = {ICML'23}
}

@article{Kusner2016PrivateCI,
  author       = {Matt J. Kusner and
                  Yu Sun and
                  Karthik Sridharan and
                  Kilian Q. Weinberger},
  editor       = {Arthur Gretton and
                  Christian C. Robert},
  title        = {Private Causal Inference},
  journal    = {International Conference on Artificial Intelligence
                  and Statistics},
  volume       = {51},
  pages        = {1308--1317},
  year         = {2016},
}

@inproceedings{Balle2018,
  author = {Balle, Borja and Wang, Yu-Xiang},
  title = {Improving the {Gaussian} Mechanism for Differential Privacy: Analytical Calibration and Optimal Denoising},
  booktitle = {Proceedings of the 35th International Conference on Machine Learning},
  series = {Proceedings of Machine Learning Research},
  volume = {80},
  pages = {394--403},
  year = {2018},
  publisher = {PMLR},
  url = {https://proceedings.mlr.press/v80/balle18a.html}
}

@article{Lee2019,
  title={Privacy-Preserving Causal Inference via Inverse Probability Weighting},
  author={Lee, Si Kai and Gresele, Luigi and Park, Mijung and Muandet, Krikamol},
  journal={arXiv preprint arXiv:1905.12592},
  year={2019}
}

@article{Li2018,
author = {Fan Li and Kari Lock Morgan and Alan M. Zaslavsky},
title = {Balancing Covariates via Propensity Score Weighting},
journal = {Journal of the American Statistical Association},
volume = {113},
number = {521},
pages = {390-400},
year  = {2018},
publisher = {Taylor & Francis},
doi = {10.1080/01621459.2016.1260466},

URL = { 
        https://doi.org/10.1080/01621459.2016.1260466
},
eprint = {     
        https://doi.org/10.1080/01621459.2016.1260466

}

}

@article{Massart1990,
 author={Pascal Massart},
 title={The Tight Constant in the {Dvoretzky--Kiefer--Wolfowitz} Inequality},
 journal={The Annals of Probability},
 year={1990}, volume={18}, number={3}, pages={1269--1283},
 doi={10.1214/aop/1176990746}
}

@article{ROSENBAUM1086,
    author = {Rosenbaum, Paul R. and Rubin, Donald B.},
    title = "{The central role of the propensity score in observational studies for causal effects}",
    journal = {Biometrika},
    volume = {70},
    number = {1},
    pages = {41-55},
    year = {1983},
    month = {04},
    issn = {0006-3444},
    doi = {10.1093/biomet/70.1.41},
    url = {https://doi.org/10.1093/biomet/70.1.41},
    eprint = {https://academic.oup.com/biomet/article-pdf/70/1/41/662954/70-1-41.pdf},
}

@inproceedings{Reimherr2019,
author = {Reimherr, Matthew and Awan, Jordan},
title = {{KNG}: The {K}-Norm Gradient Mechanism},
year = {2019},
publisher = {Curran Associates Inc.},
address = {Red Hook, NY, USA},
booktitle = {Proceedings of the 33rd International Conference on Neural Information Processing Systems},
articleno = {916},
numpages = {12}
}

@article{Rosenbaum_Rubin_1985,
author = {Paul R. Rosenbaum and Donald B. Rubin},
title = {Constructing a Control Group Using Multivariate Matched Sampling Methods That Incorporate the Propensity Score},
journal = {The American Statistician},
volume = {39},
number = {1},
pages = {33-38},
year = {1985},
publisher = {Taylor & Francis},
doi = {10.1080/00031305.1985.10479383},
URL = {https://www.tandfonline.com/doi/abs/10.1080/00031305.1985.10479383}
}

@article{Savage1971,
author = {Leonard J. Savage},
title = {Elicitation of Personal Probabilities and Expectations},
journal = {Journal of the American Statistical Association},
volume = {66},
number = {336},
pages = {783--801},
year = {1971},
publisher = {Taylor \& Francis},
doi = {10.1080/01621459.1971.10482346},
URL = { 
        https://www.tandfonline.com/doi/abs/10.1080/01621459.1971.10482346
},
eprint = { 
        https://www.tandfonline.com/doi/pdf/10.1080/01621459.1971.10482346
}
}

@article{Stuart2010,
author = {Elizabeth A. Stuart},
title = {{Matching Methods for Causal Inference: A Review and a Look Forward}},
volume = {25},
journal = {Statistical Science},
number = {1},
publisher = {Institute of Mathematical Statistics},
pages = {1 -- 21},
year = {2010},
doi = {10.1214/09-STS313},
URL = {https://doi.org/10.1214/09-STS313}
}

@article{Zhao2017,
   author = {Qingyuan Zhao and Daniel Percival},
   doi = {10.1515/jci-2016-0010},
   issn = {2193-3677},
   issue = {1},
   journal = {Journal of Causal Inference},
   month = {9},
   publisher = {Walter de Gruyter GmbH},
   title = {Entropy Balancing is Doubly Robust},
   volume = {5},
   year = {2017},
}

@article{Zhao2019,
author = {Qingyuan Zhao},
title = {{Covariate balancing propensity score by tailored loss functions}},
volume = {47},
journal = {The Annals of Statistics},
number = {2},
publisher = {Institute of Mathematical Statistics},
pages = {965 -- 993},
year = {2019},
doi = {10.1214/18-AOS1698},
URL = {https://doi.org/10.1214/18-AOS1698}
}

@article{apple2017,
  title={Learning with privacy at scale},
  author={Apple, D},
  journal={Apple Machine Learning Journal},
  volume={1},
  number={8},
  year={2017}
}

@misc{barber2014,
      title={Privacy and Statistical Risk: Formalisms and Minimax Bounds}, 
      author={Rina Foygel Barber and John C. Duchi},
      year={2014},
      eprint={1412.4451},
      archivePrefix={arXiv},
      primaryClass={math.ST},
      url={https://arxiv.org/abs/1412.4451}, 
}

@inproceedings{census2018,
author = {Abowd, John M.},
title = {The U.S. Census Bureau Adopts Differential Privacy},
year = {2018},
isbn = {9781450355520},
publisher = {Association for Computing Machinery},
address = {New York, NY, USA},
url = {https://doi.org/10.1145/3219819.3226070},
doi = {10.1145/3219819.3226070},
booktitle = {Proceedings of the 24th ACM SIGKDD International Conference on Knowledge Discovery \& Data Mining},
pages = {2867},
numpages = {1},
location = {London, United Kingdom},
series = {KDD '18}
}

@inproceedings{erlingsson2014,
  title={Rappor: Randomized aggregatable privacy-preserving ordinal response},
  author={Erlingsson, {\'U}lfar and Pihur, Vasyl and Korolova, Aleksandra},
  booktitle={Proceedings of the 2014 ACM SIGSAC conference on computer and communications security},
  pages={1054--1067},
  year={2014}
}

@article{guha2024,
 author={Sharmistha Guha and Jerome P. Reiter},
 title={Differentially private estimation of weighted average treatment effects for binary outcomes},
 journal={Computational Statistics \& Data Analysis},
 year={2025}, volume={207}, pages={108145},
 doi={10.1016/j.csda.2025.108145}
}

@article{hirano2003,
author = {Hirano, Keisuke and Imbens, Guido W. and Ridder, Geert},
title = {Efficient Estimation of Average Treatment Effects Using the Estimated Propensity Score},
journal = {Econometrica},
volume = {71},
number = {4},
pages = {1161-1189},
doi = {https://doi.org/10.1111/1468-0262.00442},
url = {https://onlinelibrary.wiley.com/doi/abs/10.1111/1468-0262.00442},
eprint = {https://onlinelibrary.wiley.com/doi/pdf/10.1111/1468-0262.00442},
year = {2003}
}

@book{imbens_rubin_2015, place={Cambridge}, title={Causal Inference for Statistics, Social, and Biomedical Sciences: An Introduction}, DOI={10.1017/CBO9781139025751}, publisher={Cambridge University Press}, author={Imbens, Guido W. and Rubin, Donald B.}, year={2015}}

@inproceedings{niu2022,
title={Differentially Private Estimation of Heterogeneous Causal Effects},
author={Fengshi Niu and Harsha Nori and Brian Quistorff and Rich Caruana and Donald Ngwe and Aadharsh Kannan},
booktitle={First Conference on Causal Learning and Reasoning},
year={2022}
}

@article{ohnishi2025,
  author  = {Yuki Ohnishi and Jordan Awan},
  title   = {Locally Private Causal Inference for Randomized Experiments},
  journal = {Journal of Machine Learning Research},
  year    = {2025},
  volume  = {26},
  number  = {14},
  pages   = {1--40},
  url     = {http://jmlr.org/papers/v26/23-1401.html}
}

@InProceedings{Awan2019,
  title = 	 {Benefits and Pitfalls of the Exponential Mechanism with Applications to {H}ilbert Spaces and Functional {PCA}},
  author =       {Awan, Jordan and Kenney, Ana and Reimherr, Matthew and Slavkovi{\'c}, Aleksandra},
  booktitle = 	 {Proceedings of the 36th International Conference on Machine Learning},
  pages = 	 {374--384},
  year = 	 {2019},
  editor = 	 {Chaudhuri, Kamalika and Salakhutdinov, Ruslan},
  volume = 	 {97},
  series = 	 {Proceedings of Machine Learning Research},
  month = 	 {09--15 Jun},
  publisher =    {PMLR},
  url = 	 {https://proceedings.mlr.press/v97/awan19a.html}
}
\endgroup

\end{document}